\documentclass{emulateapj}

\newcommand{\be}{\begin{equation}}
\newcommand{\ee}{\end{equation}}
\newcommand{\ba}{\begin{eqnarray}}
\newcommand{\ea}{\end{eqnarray}}
\newcommand{\siml}{\lower4pt \hbox{$\buildrel < \over \sim$}}
\newcommand{\simg}{\lower4pt \hbox{$\buildrel > \over \sim$}}
\def\Mesz{M\'esz\'aros}

\slugcomment{ApJ, in press}

\shorttitle{GRB early afterglows}
\shortauthors{Zhang \& Kobayashi}

\begin{document}

\title{Gamma-ray burst early afterglows: \\ reverse shock
emission from an arbitrarily magnetized ejecta}

\author{Bing Zhang\altaffilmark{1} and Shiho Kobayashi\altaffilmark{2,3}}
\affil{
$^{1}$Department of Physics, University of Nevada, Las Vegas, NV 89154
\\
$^{2}$Department of Astronomy \& Astrophysics, Pennsylvania State
University, University Park, PA 16802 \\
$^{3}$Department of Physics, Pennsylvania State University,
University Park, PA 16802}

\begin{abstract}
Evidence suggests that gamma-ray burst (GRB) ejecta are likely
magnetized, although the degree of magnetization is unknown. When such
magnetized ejecta are decelerated by the ambient medium, the
characteristics of the reverse shock emission are strongly influenced
by the degree of magnetization. We derive a rigorous analytical
solution for the relativistic 90$^{\rm o}$ shocks under the ideal MHD
condition. The solution is reduced to the
Blandford-McKee hydrodynamical solution when the magnetization
parameter $\sigma$ approaches zero, and to the Kennel-Coroniti
solution (which depends on $\sigma$ only) when the shock
upstream and downstream are ultra-relativistic with each other. Our
generalized solution can be used to treat the more general cases,
e.g. when the shock upstream and downstream are mildly
relativistic with each other. We find that the suppression factor of
the shock in the strong magnetic field regime is only mild as
long as the shock upstream is relativistic with respect to the
downstream, and it saturates in the high-$\sigma$ regime. This
indicates that generally strong relativistic shocks still exist in the
high-$\sigma$ limit. This can effectively convert 
kinetic energy into heat. The overall efficiency of converting ejecta
energy into heat, however, decreases with increasing $\sigma$, mainly
because the fraction of the kinetic energy in the total energy
decreases. We use the theory to study the reverse shock emission
properties of arbitrarily magnetized ejecta in the GRB problem
assuming a constant density of the circumburst medium. We 
study the shell-medium interaction in detail and categorize various
critical radii for shell evolution. With typical GRB parameters, a
reverse shock exists when $\sigma$ is less than a few tens or a few
hundreds. The shell evolution can be still
categorized into the thick and thin shell regimes, but the separation
between the two regimes now depends on the $\sigma$ parameter and
the thick shell regime greatly shrinks at high-$\sigma$. The thin
shell regime can be also categorized into two sub-regions depending 
on whether the shell starts to spread during the first shock
crossing. The early optical afterglow lightcurves are
calculated for GRBs with a wide range of $\sigma$ value, with the main
focus on the reverse shock component. We find 
that as $\sigma$ increases from below the reverse
shock emission level increases steadily until reaching a peak at
$\sigma \siml 1$, then it decreases steadily when $\sigma > 1$. 
At large $\sigma$ values, the
reverse shock peak is broadened in the thin shell regime because of the
separation of the shock crossing radius and the deceleration
radius. This novel feature can be regarded as a signature of high
$\sigma$.  The early afterglow data of GRB 990123 and GRB 021211 could
be understood within the theoretical framework developed in this
paper, with the inferred $\sigma$ value $\simg 0.1$. The case of GRB
021004 and GRB 030418 may be also interpreted with higher $\sigma$
values, although more detailed modeling is needed. Early tight optical
upper limits could be interpreted as very high $\sigma$ cases, in
which a reverse shock does not exist or very weak. 
Our model predictions could be further tested
against future abundant early afterglow data collected by the Swift
UV-optical telescope, so that the magnetic content of GRB fireballs
can be diagnosed. 
\end{abstract}


\section{Introduction}

Extensive broad-band observational campaigns and theoretical
modeling of gamma-ray burst (GRB) afterglows have greatly advanced
our understanding of these mysterious cosmic explosions. Yet, the
origin of the GRB prompt emission itself and the nature of the
relativistic flow (which are directly connected to the function of the
central engine) is still unknown (e.g. \Mesz~2002; Zhang \&
\Mesz~2004). In particular, it is unclear how important the role
of magnetic fields is in producing GRBs. Recently two
independent pieces of evidence suggest that the GRB central engine
is likely strongly magnetized. First, the claimed detection of the
very high degree of linear polarization of gamma-ray emission in GRB 
021206 (Coburn \& Boggs 2003, see however Rutledge \& Fox 2004), if
true, could be readily interpreted by assuming 
that the magnetic field involved in the synchrotron
radiation is globally ordered (e.g. Waxman 2003;
Granot 2003), although some alternative explanations remain (e.g.
Waxman 2003). Second, recently we (Zhang, Kobayashi \& \Mesz~2003,
hereafter ZKM03)
developed a method to perform combined reverse and forward shock
emission study for GRB early optical afterglows, and revealed that
a stronger magnetic field in the reverse shock region than in the
forward shock region is needed to interpret the early afterglow data
of GRB 990123 and GRB 021211. This claim was confirmed by independent
detailed case studies for both bursts (Fan et al. 2002; Kumar \&
Panaitescu 2003). These findings suggest that magnetic fields may
play a significant role in the GRB physics, as has been suggested
by various authors previously (e.g. Usov 1994; Thompson 1994;
\Mesz~\& Rees 1997b; Wheeler et al. 2000; Spruit, Daigne \&
Drenkhahn 2001; Blandford 2002). Within the framework of the
currently favored collapsar progenitor model for GRBs (MacFadyen \&
Woosley 1999), the ejecta are found to be magnetized when MHD
simulations are performed (Proga et al. 2003).

The degree of magnetization of the ejecta, however, is unknown. This
is usually quantified by the parameter $\sigma$ (see eq.[\ref{sigma}]
for a precise definition), the ratio of the electromagnetic energy flux
to the kinetic energy flux. Current GRB models are focused on two
extreme regimes. In the first regime, it is essentially assumed that
the GRB fireball is purely hydrodynamical. Magnetic fields are
introduced only through an equipartition parameter $\epsilon_B$ (which
is of the order of 0.001-0.1) for the purpose of calculating synchrotron
radiation. This is the $\sigma \rightarrow 0$ regime. In this picture,
the GRB prompt emission is produced from internal shocks (Rees \&
\Mesz~1994) or sometimes from external shocks (\Mesz~\& Rees 1993;
Dermer \& Mitman 1999). This is currently the standard scenario of
GRB emission. The second is the $\sigma \rightarrow \infty$
regime. This is the regime where a Poynting-flux dominates the flow,
and GRB prompt emission is envisaged to be due to some less
familiar magnetic dissipation processes (e.g. Usov 1994; Spruit et
al. 2001; Blandford 2002; Lyutikov \& Blandford 2003). In principle, a
GRB event could include both a ``hot component'' as invoked in the
$\sigma=0$ model (e.g. due to neutrino annihilation) and a ``cold
component'' as invoked in the $\sigma=\infty$ model, the interplay
between both components may result in a $\sigma$ value varying in a
wide range (Zhang \& \Mesz~2002). It is an important but difficult
task to pin down the degree of magnetization of GRB ejecta.

GRB early afterglow data (especially in the optical band)
potentially contain essential information to diagnose the magnetic
content of the fireball. The reason is that an early optical afterglow
lightcurve is believed to include contributions from  both
the forward shock (which propagates into the ambient medium) and
the reverse shock (which propagates into the ejecta itself). Since
the magnetization degree of the ejecta influences the emission 
level of the reverse shock (or maybe even the level of the forward
shock), by studying the 
interplay between the reverse shock and the forward shock emission
components, one could potentially infer the degree of magnetization of
the ejecta. In all the current analyses, the reverse shock
emission is treated purely hydrodynamically (e.g. \Mesz~\& Rees
1997a; Sari \& Piran 1999; Kobayashi 2000; Kobayashi \& Zhang
2003a; ZKM03). When confronted with the available early afterglow
data (four cases so far: GRB 990123, Akerlof et al. 1999; 
GRB 021004, Fox et al. 2003a; GRB 021211, Fox et al. 2003b, Li et
al. 2003a; and GRB 030418, Rykoff et al. 2004), the model works
reasonably well for two of them (GRB 990123 and GRB 021211), although
a good fit requires that the magnetic field in the reverse shock
region is much stronger than that in the forward shock region
(ZKM03). For the other two, the lightcurves are not easy to explain
with the simplest reverse shock model. On the other hand, GRB
ejecta could in principle have an arbitrary $\sigma$ value. When
$\sigma$ is large, the conventional hydrodynamical treatment is no
longer a good approximation, and a full treatment involving MHD shock 
jump condition is desirable.

It is generally believed that a GRB involves a rapidly-rotating
central engine. If the magnetic dissipation processes are not
significant, field lines are essentially frozen in the expanding
shells. The radial component of the magnetic field decays with
radius as $\propto R^{-2}$, while the toroidal magnetic field
decays as $\sim R^{-1}$. At the external shock
radius, magnetic field lines are essentially
frozen in the plane perpendicular to the moving direction. The MHD
shock Rankine-Hugoniot relations are greatly simplified in such a
90$^{\rm o}$ shock. Such relations have been studied extensively
before both analytically and numerically. Kennel \& Coroniti
(1984) derived some simplified analytical expressions applicable
for strong 90$^{\rm o}$ degree shocks whose upstream and downstream
are ultra-relativistic with each other. The model
was used to treat the pulsar-wind nebula problem. In this
regime, the strength of the shock is essentially characterized by only
one parameter, i.e. the $\sigma$ parameter. The conclusion was
confirmed later by numerical simulations (e.g. Gallant et al. 1992)

Within the context of GRBs, since a GRB invokes a transient
release of energy, the ejecta shell has a finite width (in
contrast to the long-standing pulsar wind). Under some conditions, 
the reverse shock upstream and downstream could never become
relativistic with each other when the reverse shock crosses the 
ejecta shell.
In the $\sigma=0$ limit, whether the reverse shock becomes
relativistic depends on the comparison between the time scale ($T$) of
the central engine activity (essentially the duration of the burst)
and the time scale ($t_\gamma$) when the mass of the ambient medium
collected by the fireball reaches $1/\gamma_0$ times the mass of the
ejecta (e.g. Sari \& Piran 1995; Kobayashi, Piran
\& Sari 1999). Both times are measured by the observer. The case of 
$T>t_\gamma$ is called the thick shell regime, and the reverse shock
is relativistic. In many cases, however, one has
$T<t_\gamma$. i.e. the thin shell regime\footnote{If a GRB contains
several well-separated emission episodes, the whole burst may be even
separated into several discrete shells. In such cases, even a
long-duration burst may be treated as the superposition of several
thin shells rather than one single thick shell. See Zhang \&
\Mesz~2004 for more discussions.}. The reverse shock is initially
non-relativistic and only becomes mildly relativistic as the shock
crosses the shell. For magnetized ejecta (e.g. a shell with a finite
width but an arbitrary $\sigma$ value), the separation between the
thick and thin shell regimes becomes more complicated, but the
non-relativistic reverse shock case (for the thin shell regime) is
even more common (see \S\ref{sec:regimes}). The Kennel-Coroniti
approximation can not be directly used. The theory developed in this
paper becomes essential to discuss the reverse shock physics in this
parameter regime. 

In this paper, we present a detailed treatment of reverse
shock emission for an arbitrarily magnetized ejecta under the ideal
MHD condition. The reverse
shock emission in the mildly magnetized regime is also discussed
by Fan, Wei \& Wang (2004a) recently. Here we will develop a
theoretical framework to include discussions for the reverse shock
emission in a wider $\sigma$ range, as well as for various ejecta-medium
interaction parameter regimes. We first (\S2) present a rigorous
analytical solution for the MHD 90$^{\rm o}$ shock jump
conditions, which is applicable for an arbitrary $\sigma$ value
and for an arbitrary Lorentz factor $\gamma_{21}$ between the
upstream and the downstream. The detailed derivation and the relevant
equations are presented in the Appendix A. We then (\S3) discuss the
ejecta - medium interaction within the context of GRB fireball
deceleration and re-investigate the critical fireball radii and
re-categorize the thick vs. thin shell regimes. This leads to a more
complicated picture than in the pure hydrodynamical case (Sari \&
Piran 1995). In \S4, we calculate the synchrotron emission from
the shocks under the conventional assumptions
about the particle acceleration in collisionless shocks, and present
the predicted GRB early optical lightcurves for a wide range of
$\sigma$ value. We discuss how the early afterglow data may be used to
diagnose the magnetic content of GRB ejecta. Our results are
summarized in \S5 with some discussions. 

\section{Analytical solution of the relativistic 90$^{\rm o}$ shocks
\label{sec:shock}}

We now consider a relativistic shock that propagates into a magnetized 
medium. In the following
analysis, the unshocked region (upstream) is denoted as region 1,
the shocked region (downstream) is denoted as region 2, and the
shock itself is denoted as ``$s$''\footnote{Notice that such a notation
system is only valid for \S\ref{sec:shock} and the Appendix A. When
discussing the GRB problem, i.e. the ejecta-medium interaction
(\S\ref{sec:dyn}), we introduce different meanings for the subscript
numbers.}. Hereafter $Q_{ij}$ denotes the value 
of the quantity $Q$ in the region ``$i$'' in the rest frame of ``$j$'',
and $Q_i$ denotes the value of the quantity $Q$ in the region ``$i$'' in
its own rest frame. For example, $\gamma_{12}$ is the relative Lorentz
factor between the regions 1 and 2, $\beta_{1s}$ is the relative
velocity (in unit of the speed of light $c$) between the region 1 and
the shock, $B_{2s}$ is the magnetic field strength of the region 2
in the rest frame of the shock, while $B_1$ is the comoving magnetic
field strength in the region 1, etc. The relativistic 90$^{\rm o}$
shock Rankine-Hugoniot relations could be written as (Kennel \& Coroniti
1984)
\begin{eqnarray}
n_1 u_{1s}        & = & n_2 u_{2s}~, \label{jump1}\\
{\cal E}= \beta_{1s} B_{1s} & = & \beta_{2s} B_{2s}~, \label{EB} \\
\gamma_{1s} \mu_1 + \frac{{\cal E} B_{1s}}{4\pi n_1 u_{1s}} & = &
  \gamma_{2s} \mu_2 + \frac{{\cal E} B_{2s}}{4\pi n_2 u_{2s}}~, \label{jump2}\\
\mu_1 u_{1s}+\frac{p_1}{n_1 u_{1s}} + \frac{B_{1s}^2}{8\pi n_1u_{1s}}
  & = & \mu_2 u_{2s}+\frac{p_2}{n_2 u_{2s}}
  + \frac{B_{2s}^2}{8\pi n_2 u_{2s}}~, \label{jump3}
\end{eqnarray}
where $\beta$ denotes the dimensionless velocity,
$\gamma=(1-\beta^2)^{-1/2}$ denotes the Lorentz factor, and
$u=\beta \gamma$ denotes the radial four velocity, so that
$\gamma^2=1+u^2$. Hereafter, $n$, $e$, $p=(\hat\Gamma-1)
e$ denote the number density, internal energy and
thermal pressure, respectively, and $\hat\Gamma$ is the adiabatic
index. The enthalpy is $n m_p c^2+e+p$, and the specific enthalpy
can be written as
\be
\mu = m_p c^2 +
\frac{\hat\Gamma}{\hat\Gamma-1}\left(\frac{p}{n}
\right),\label{mu}
\ee
where $m_p$ is the proton mass and $c$ is
the speed of light. It is convenient to define a parameter
\be
\sigma_i=\frac{B_i^2}{4\pi n_i\mu_i}=\frac{B_{is}^2}{4\pi n_i\mu_i
\gamma_{is}^2},
\ee
to denote the degree of magnetization in each
region. The magnetization parameter in the upstream region
($\sigma_1$) is a more fundamental parameter, since it
characterizes the magnetization of the flow itself. We therefore
define\footnote{Notice that the definition is slightly different
from that in Kennel \& Coroniti (1984). We find that our
definition allows the parameters to be coasted into an analytical
form as a function of $\sigma$ and $\gamma_{21}$.}
\be
\sigma
\equiv \sigma_1 = \frac{B_{1s}^2}{4\pi n_1\mu_1\gamma_{1s}^2}.
\label{sigma}
\ee
In our problem, we are interested in a ``cold''
upstream flow, i.e., $e_1=p_1=0$, so that $\mu_1=m_p c^2$. This is
the only assumption made in the derivation. After some algebra (see
Appendix), we can finally write 
\be
\frac{e_2}{n_2 m_p c^2} = (\gamma_{21}-1)\left[1
-\frac{\gamma_{21}+1} {2 u_{1s}(\gamma_{21},\sigma)
u_{2s}(\gamma_{21},\sigma)}\sigma\right].
\label{e/n}
\ee
Here $u_{2s}(\gamma_{21},\sigma)$ is a function of $\gamma_{21}$ and
$\sigma$, and can be solved once $\gamma_{21}$ and $\sigma$ are known.
After some analytical treatments of the relativistic Rankine-Hugoniot
relations (eq.[\ref{jump1}-\ref{jump3}]), one can come up with an
equation to solve $u_{2s}$. Since it is complicated, we only present
it in the Appendix (eq.[\ref{eq}]). Once $u_{2s}$ is solved, we can
also solve $u_{1s}$ (using eq.[\ref{LT}]), i.e.,
\be
u_{1s}(\gamma_{21},\sigma)=u_{2s}(\gamma_{21},\sigma) \gamma_{21}
+[u_{2s}^2(\gamma_{21},\sigma)+1]^{1/2} (\gamma_{21}^2-1)^{1/2}~.
\label{u1s}
\ee
The compressive ratio can be derived directly from eq.(\ref{jump1}),
i.e. 
\be
\frac{n_2}{n_1} = \frac{u_{1s}(\gamma_{21},\sigma)}
{u_{2s}(\gamma_{21},\sigma)} = \gamma_{21}
+\frac{[u_{2s}^2(\gamma_{21},\sigma)+1]^{1/2}}
{u_{2s}(\gamma_{21},\sigma)} (\gamma_{21}^2-1)^{1/2}~. \label{n2/n1} 
\ee
The main point here is that both $e_2/n_2 m_p c^2$ and $n_2/n_1$ can be
determined by two unknown parameters, i.e. $\gamma_{21}$ and $\sigma$,
so that when they are given arbitrarily, the whole problem is solved.

In the downstream region, the total pressure includes the contribution 
from the comoving thermal pressure $p_2=(\hat\Gamma-1) e_2$ and the
comoving magnetic pressure $p_{b,2}=B_{2}^2/8\pi$. The ratio between
the magnetic pressure to the thermal pressure is also a function of
$\sigma$ and $\gamma_{21}$: 
\ba
\frac{p_{b,2}}{p_2} & = & \left(\frac{\beta_{1s}}{\beta_{2s}}\right)^2
\left(\frac{4\pi n_1 m_p c^2 \gamma_{1s}^2 \sigma}
{8\pi \gamma_{2s}^2 (\hat\Gamma-1) e_2}\right) \nonumber \\
& = & \frac{1}{2(\hat\Gamma-1)}\left(\frac{u_{1s}}{u_{2s}}\right)
\sigma \left(\frac{e_2}{n_2 m_p c^2}\right)^{-1},
\label{Rp}
\ea
where eqs.(\ref{EB}) and (\ref{sigma}) have been used.

The correctness of the solution (eq.[\ref{eq}]) are verified in two
asymptotic regimes.

\subsection{The $\sigma=0$ regime \label{sec:sigma=0}}

When $\sigma=0$, the equation to solve
$u_{2s}^2(\gamma_{21},\sigma)$ (eq.[\ref{eq}]) is greatly
simplified (eq.[\ref{eq1}]). All quantities can be expressed as a
function of $\gamma_{21}$. The solutions are 

\ba
u_{2s}^2 & = & \frac{(\gamma_{21}-1)(\hat\Gamma -1)^2}
{\hat\Gamma (2-\hat\Gamma)(\gamma_{21}-1)+2} \label{bm0} \\
u_{1s}^2 & = & \frac{(\gamma_{21}-1)(\hat\Gamma \gamma_{21}+1)^2}
{\hat\Gamma (2-\hat\Gamma)(\gamma_{21}-1)+2} \\
\frac{e_2}{n_2} & = & (\gamma_{21}-1) m_pc^2 \label{bm1}\\
\frac{n_2}{n_1} & = & \frac{\hat\Gamma\gamma_{21}+1}{\hat\Gamma-1}
\label{bm2} \\
\gamma_{1s}^2 & = & \frac{(\gamma_{21}+1)[\hat\Gamma
(\gamma_{21}-1)+1]^2}{\hat\Gamma(2-\hat\Gamma)(\gamma_{21}-1) +2}
\label{bm3} \ea The equations (\ref{bm1}-\ref{bm3}) are just
equations (3-5) of Blandford \& McKee (1976), and they are the
starting point for the hydrodynamical analysis of the reverse
shock emission (e.g. Sari \& Piran 1995). Under the limit of
$\gamma_{21} \gg 1$ and $\hat\Gamma=4/3$ (i.e. the downstream fluid 
is relativistic), the equations (\ref{bm2}) and
(\ref{bm3}) are reduced to the familiar forms of
$n_2/n_1=4\gamma_{21}+3$, $\gamma_{1s} \simeq \sqrt{2}
\gamma_{21}$ and $\gamma_{2s} \simeq 3\sqrt{2}/4$ (or
$u_{2s}\simeq \sqrt{2}/4$).

\subsection{The $\gamma_{21} \gg 1$ regime \label{sec:gam>>1}}

In the $\gamma_{21} \rightarrow \infty$ limit, the equation for
$u_{2s}^2(\gamma_{21},\sigma)$ (eq.[\ref{eq}]) is also
simplified (eq.[\ref{eq2}]). The solution
of $u_{2s}$ is a function of $\sigma$ only, which reads
\begin{equation}
u_{2s}^2 = \frac{\hat\Gamma(1-\frac{\hat\Gamma}{4})\sigma^2
+(\hat\Gamma^2-2\hat\Gamma+2)\sigma+(\hat\Gamma-1)^2 + \sqrt{X}}
{2\hat\Gamma(2-\hat\Gamma)(\sigma+1)} \label{u2skc}
\end{equation}
where
\ba
X & = & \hat\Gamma^2(1-\frac{\hat\Gamma}{4})^2\sigma^4 + \hat\Gamma
(\frac{\hat\Gamma^3}{2}-3\hat\Gamma^2+7\hat\Gamma-4)\sigma^3 \nonumber
\\
& + &
(\frac{3}{2}\hat\Gamma^4-7\hat\Gamma^3+\frac{31}{2}\hat\Gamma^2
-14\hat\Gamma+4)\sigma^2 \nonumber \\ 
& + & 2(\hat\Gamma-1)^2
(\hat\Gamma^2-2\hat\Gamma+2)\sigma+(\hat\Gamma-1)^4.
\ea
Notice that there are two solutions with the term $\pm\sqrt{X}$ in
the numerator of eq.(\ref{u2skc}), but the minus term leads to
negative pressure and is un-physical. For a relativistic downstream
region, i.e. $\hat\Gamma=4/3$, the solution is reduced to
\ba
u_{2s}^2 & = & \frac{8\sigma^2+10\sigma+1+\sqrt{64\sigma^2(\sigma+1)^2
+20\sigma(\sigma+1)+1}}{16(\sigma+1)} \nonumber \\
& = & \frac{8\sigma^2+10\sigma+1+(2 \sigma+1)
\sqrt{16\sigma^2+16\sigma+1}}{16(\sigma+1)}  
\label{u2skc2}
\ea
This is the eq.(4.11) of Kennel \& Coroniti (1984). With $u_{2s}$,
one can derive $u_{1s}$ using equation (\ref{u1s}), which depends on 
$\gamma_{21}$ as well. The quantities $e_2/n_2 m_p c^2$ and $n_2/n_1$
can be also derived accordingly. In the $\sigma=0$ limit, equation
(\ref{u2skc2}) is reduced to $u_{2s} \simeq \sqrt{2}/4$, which is
consistent with the asymptotic results in \S\ref{sec:sigma=0}.

\subsection{The general cases \label{sec:general}}

For more general cases with arbitrary values of $\gamma_{21}$
and $\sigma$, $u_{2s}^2(\gamma_{21},\sigma)$ has to be solved
rigorously. The equation (eq.[\ref{eq}]) is solved numerically, and
the solutions indeed show deviations from both asymptotic regimes for
arbitrary $\gamma_{21}$ and $\sigma$ values.

\begin{figure}
\epsscale{.80}
\plotone{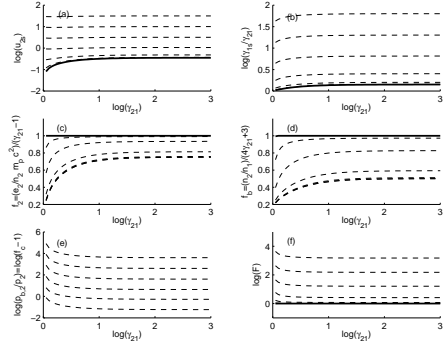}
\caption{
The variations of six parameters, i.e., $u_{2s}$,
$\gamma_{1s}/\gamma_{21}$, $e_2/n_2 m_p c^2$, $n_2/n_1$,
$p_{b,2}/p_2$, and $F$, as a function of $\gamma_{21}$. The thick
solid line indicates the case for $\sigma=0$, which is the 
Blandford-McKee (1976) solution. For $p_{b,2}/p_2$ (panel e), the $\sigma=0$
line is at negative infinity. The dashed lines, starting from the one
closest to the thick line, are for $\sigma=0.01, 0.1, 1, 10, 100,
1000$, respectively. The parameter $e_2 / n_2 m_p c^2$ (random Lorentz
factor in the shocked, downstream region) is normalized to
$(\gamma_{21}-1)$, and the parameter $n_2/n_1$ (compressive ratio) is
normalized to $(4\gamma_{21}+3)$, both are the values expected in the
$\sigma=0$ case. 
\label{fig:gam}}
\end{figure}

\begin{figure}
\epsscale{.80}
\plotone{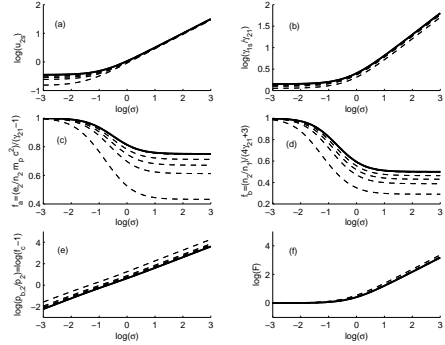}
\caption{
The variations of six parameters, i.e., $u_{2s}$,
$\gamma_{1s}/\gamma_{21}$, $e_2/n_2 m_p c^2$, $n_2/n_1$, 
$p_{b,2}/p_2$, and $F$, as a function of $\sigma$. The thick solid
line is the Kennel-Coroniti (1984) solution, denoting a
$\gamma_{21} \gg 1$ regime. The 
dashed lines, starting from the one closest to the thick line, are for 
$\gamma_{21}=1000, 100, 10, 5, 3, 1.5$, respectively. Again the
parameters $e_2 / n_2 m_p c^2$ and $n_2/n_1$ are normalized to 
$(\gamma_{21}-1)$ and $(4\gamma_{21}+3)$, respectively.
\label{fig:sig}}
\end{figure}

Figure 1 shows the variations of six parameters, i.e., $u_{2s}$,
$\gamma_{1s}/\gamma_{21}$, $e_2/n_2 m_p c^2$, $n_2/n_1$,
$p_{b,2}/p_2$, and $F$ (see definition in eq.[\ref{F}]) as a
function of $\gamma_{21}$. The thick solid line indicates the case for 
$\sigma=0$, which is the strict Blandford-McKee (1976) regime. The
dashed lines, starting from the one closest to the thick line, are for 
$\sigma=0.01, 0.1, 1, 10, 100, 1000$, respectively. In order to find
out the correction factors to the pure hydrodynamical case, we
normalize $e_2/n_2 m_p c^2$ and $n_2/n_1$ with respect to the
$\sigma=0$ case. $\hat\Gamma=4/3$ has been adopted. We find that all
the parameters achieve asymptotic values when $\gamma_{21} \gg 1$, and
that the asymptotic value depends on the 
value of $\sigma$. In Figure 2 we plot the variations of the
same six parameters as a function of $\sigma$. The thick
solid line is the $\gamma_{21} \gg 1$ Kennel-Coroniti (1984) limit,
and the other dashed lines, starting from the closest to the thick
line, correspond to $\gamma_{21} = 1000, 100, 10, 5, 3, 1.5$,
respectively. For $\gamma_{21} > 100$, the Kennel-Coroniti
approximation is good enough.

An obvious conclusion from Figs. 1 and 2 is that all the (normalized)
parameters are insensitive to $\gamma_{21}$ (especially when
$\gamma_{21} >$ a few), but are sensitive to $\sigma$. Both $u_{2s}$
and $\gamma_{1s}/\gamma_{21}$ increase with $\sigma$, while both
$e_2/n_2 m_p c^2$ and $n_2/n_1$ decrease with $\sigma$. For
$\gamma_{1s}/\gamma_{21}$, as long as $\gamma_{21}$ is mildly large
(e.g. $>3$), the ratio is essentially the function of $\sigma$
only. It starts from the 
conventional value $\sqrt{2}$ in the $\sigma \sim 0$ regime and
increases quickly as $\sigma$ approaches unity, which means that the
shock leads the fluid substantially (in the upstream rest frame) in
the high $\sigma$ regime. Both $e_2/n_2 m_p c^2$ and $n_2/n_1$ are
suppressed 
when $\sigma$ increases, but the suppression factor (with respect to
the $\sigma=0$ limit) is not large, especially when $\gamma_{21}$ is
not too small. For example, for $\gamma_{21}>3$, the suppression
factor for $e_2/n_2$ is larger than 0.6, while that for $n_2/n_1$ is
larger than 0.4. Furthermore, the suppression factor reaches an
asymptotic value when $\sigma$ approaches several. This result is very 
interesting, since conventionally it is believed that the shock is
completely suppressed when $\sigma$ reaches larger values.
Our results suggest that {\em relativistic strong shocks still exist
in the high-$\sigma$ regime.} The suppression factor, which
essentially does not depend on $\sigma$, is only mild as long as the
shock is relativistic. The overall efficiency of converting the total
energy (the kinetic energy plus the Poynting flux energy) to heat still
decreases steadily with increasing $\sigma$. The reason is not that
the shock (which converts the kinetic energy into heat) itself is less
strong, but is that the fraction of the kinetic energy in the total
energy, i.e., $(1+\sigma)^{-1}$, becomes smaller as $\sigma$ becomes
larger. 

\section{Ejecta - medium interaction\label{sec:dyn}}

\subsection{Basic equations}

Now we consider an arbitrarily magnetized flow with 
magnetization parameter $\sigma$ and Lorentz factor 
$\gamma=\gamma_4$ being decelerated by an ambient medium with density
$n=n_1$. A pair of shocks form when the shock forming condition is
satisfied, i.e. the relative velocity between the two colliders exceeds
the sound velocity in the medium and the magnetoacoustic wave velocity of
the ejecta, and that the pressure in the shocked region exceeds the
pressure in the unshocked region. In the GRB case (a relativistic
ejecta), a forward shock always form, while a reverse shock may not
always form if $\sigma$ is too large. In the high-$\sigma$ regime, the
magnetoacoustic wave velocity is essentially the 
Alfven velocity for a $90^{\rm o}$ shock. The first condition for the
reverse shock formation is $\gamma_{41} \gg \gamma_{A} \sim
\sqrt{1+\sigma}$, where $\gamma_{A}$ is the Alfven Lorentz factor in
the ejecta. For GRBs we have $\gamma_{41} \geq 100$ (to ensure that
the observed gamma-ray spectrum is non-thermal), so that this
condition is satisfied as long as $\sigma \leq 10^4$. The second
condition is generally more stringent, which is expressed in
eqs. (\ref{RS-condition-1}) and (\ref{RS-condition-2}) below. In any
case, with reasonable parameters a reverse shock could be formed when
$\sigma$ is less than hundreds or tens. When the reverse shock forms,
we can then investigate a picture where
two shocks and one contact discontinuity separate the ejecta
and medium into four regions. Below we take the usual convention to
define the four regions: (1) unshocked medium; (2) shocked medium; (3)
shocked ejecta; (4) unshocked ejecta. Notice that hereafter the
numerical subscripts have different meanings from the ones used
in \S\ref{sec:shock} and Appendix, where the numbers ``1'' and ``2''
denote upstream and downstream, respectively. For the forward shock,
both sets of notations coincident, while for the reverse shock, the
previous ``1'' and ``2'' are replaced by ``4'' and ``3'',
respectively. We assume that the ambient medium is not magnetized
so that $\sigma_1=0$, but will assign an arbitrary magnetization
parameter 
\be
\sigma\equiv \sigma_4 = \frac{B_4^2}{4\pi n_4 m_p c^2} =
\frac{B_{4s}^2}{4\pi n_4 m_p c^2 \gamma_{4s}^2} 
\label{sigma4}
\ee
for the ejecta. Since we are discussing the problem in the rest frame
of the medium, we will drop out the subscript ``1'' whenever it means
``in the rest frame of the region 1''. We can then write the following 
relations based on the shock jump conditions. Throughout the following
discussions, $\hat\Gamma=4/3$ is adopted.
\ba
\frac{e_2}{n_2 m_p c^2} & = & (\gamma_2-1) \simeq \gamma_2, \label{sp1} \\
\frac{n_2}{n_1} & = & 4\gamma_2 + 3 \simeq 4 \gamma_2, \label{sp2} \\
\frac{e_3}{n_3 m_p c^2} & = & (\gamma_{34}-1) f_{a}, \label{sp3} \\
\frac{n_3}{n_4} & = & (4\gamma_{34}+3) f_b ~, \label{sp4}
\ea
where
\ba
f_a & = & f_a(\sigma,\gamma_{34}) \nonumber \\
& = &1 - \frac{\gamma_{34}+1}
{2[u_{3s}^2\gamma_{34}
+u_{3s}(u_{3s}^2+1)^{1/2}(\gamma_{34}^2-1)^{1/2}]}  
 \sigma 
\label{fa}
\ea
and 
\be
f_b=f_b(\sigma,\gamma_{34}) = \frac{ \gamma_{34}+ \frac{(u_{3s}^2+1)^{1/2}}
 {u_{3s}}(\gamma_{34}^2-1)^{1/2} }{4 \gamma_{43}+3}
\label{fb}
\ee
are the correction factors for $e_3/n_3 m_p c^2$ and $n_3/n_4$ with
respect to the $\sigma=0$ limit, and $u_{3s}$ is a function of
$\gamma_{34}$ and $\sigma$, whose solution is found in the Appendix A
(eq.[\ref{eq}]). The functions $f_a$ and $f_b$ are plotted as a
function of $\sigma$ in Fig.(\ref{fig:sig}c) and Fig.(\ref{fig:sig}d),
respectively, which shows $f_a \rightarrow 1$ and $f_b \rightarrow 1$ 
when $\sigma \rightarrow 0$. Constant speed across the contact
discontinuity gives 
\be
\gamma_2=\gamma_3~,
\label{gam2=gam3}
\ee
and constant pressure across the contact discontinuity gives
\be
\frac{e_2}{3} = \frac{e_3}{3} + \frac{B_{3}^2}{8\pi} =
\frac{e_3}{3} (1+\frac{p_{b,3}}{p_3})~,
\ee
or 
\be
e_2=e_3 f_c~,
\label{p2=p3}
\ee
where
\ba
f_c=f_c(\sigma,\gamma_{34})=1+\frac{p_{b,3}}{p_3}, & (\propto \sigma)
\label{fc}
\ea
The pressure ratio $p_{b,3}/p_3$ is calculated according to
eq.(\ref{Rp}), whose dependence on $\sigma$ is plotted in
Fig.(\ref{fig:sig}e) which shows a $\propto \sigma$ dependence in the
$\sigma \gg 1$ asymptotic regime. Hereafter we denote the asymptotic
behavior in the $\sigma \gg 1$ limit in a pair of
parenthesis immediately following an equation. 

In order to have equation (\ref{p2=p3}) satisfied, the condition
should be that the thermal pressure generated in the forward shock is
stronger than the magnetic pressure in the region 4. This gives
$B_4^2/8\pi<(4/3)\gamma_4^2 n_1 m_p c^2$. Noticing equation
(\ref{sigma4}), the condition for the existence of the reverse shock
can be written as
\be
\sigma < \frac{8}{3} \gamma_4^2 \frac{n_1}{n_4}~.
\label{RS-condition-1}
\ee

When the reverse shock exists, using equations (\ref{sp1}-\ref{sp4})
and (\ref{p2=p3}), one can finally get
\be 
\frac{n_4}{n_1} F=\frac{
(\gamma_2-1)(4\gamma_2+3)} {(\gamma_{34}-1)(4\gamma_{34}+3)}~,
\label{rel}
\ee
where
\be
F = f_a f_b f_c~,
\label{F}
\ee
and $f_a$, $f_b$ and $f_c$ are defined in equations (\ref{fa}),
(\ref{fb}) and (\ref{fc}), respectively. The parameter $F$ has been
calculated for different input parameters, and the results are shown
in Fig.(1f) and Fig.(2f), respectively. We can see that $F$ is very
insensitive to $\gamma_{34}$, and is essentially a function of
$\sigma$ only. The asymptotic behavior in the $\sigma \gg 1$ regime
is $F(\sigma) \propto \sigma$. We then write 
\ba
F(\gamma_{34},\sigma) \simeq F(\sigma), & (\propto \sigma).
\ea

The equation (\ref{rel}) can be used to define whether the reverse
shock upstream and downstream are relativistic with each
other. Similar to the analysis of Sari \& Piran 
(1995), we analyze the value of $(n_4/n_1) F / \gamma_4^2$. 
The relative Lorentz factor of the reverse shock upstream and
downstream is 
\be
\gamma_{34} \simeq \frac{1}{2} \left(\frac{\gamma_4}{\gamma_3}
+\frac{\gamma_3}{\gamma_4}\right).
\label{gam34}
\ee
For the relativistic case, i.e. $\gamma_{34} \gg 1$, we
have $(n_4/n_1) F / \gamma_4^2 \sim \gamma_2^2/(\gamma_{34}^2
\gamma_4^2) \sim \gamma_2^4/\gamma_4^4 \ll 1$. On the other hand, for
a non-relativistic case, we have $\gamma_{34} \sim 1$,
$\gamma_4 \sim \gamma_3$ and $(\gamma_{34}-1)(4 \gamma_{34}+3)
=\epsilon \ll 1$. This gives $(n_4/n_1) F / \gamma_4^2 \sim 1/\epsilon
\gg 1$. We thus conclude that the reverse shock upstream is
relativistic with respect to the downstream when
$\gamma_4^2 \gg (n_4/n_1) F$, while it is non-relativistic when 
$\gamma_4^2 \ll (n_4/n_1) F$. For $\sigma=0$, we have $F=1$, and the
result is fully consistent with Sari \& Piran (1995).

\subsection{Critical radii}

We consider an isotropic fireball with total energy $E=E_K+E_P$, where 
$E_K$ is the kinetic energy and $E_P$ is the Poynting flux
energy. The discussions are also valid for a collimated jet by 
regarding the various energy components as the ``isotropic'' 
values. With the 
definition of $\sigma$ (eq.[\ref{sigma}]), we find $E_P/E_K \sim
\sigma$ (from eq.[\ref{jump3}])\footnote{The full presentation of
eq.[\ref{jump3}] should be $\mu_1 u_{1s} + p_1/n_1u_{1s} + B^2_{1s} /
8\pi n_1 u_{1s}+B_{1s}^2 \beta_{1s}^2/8\pi n_1u_{1s} = \mu_2 u_{2s} +
p_2/n_2u_{2s}+B_{2s}^2/8\pi n_2 u_{2s}+B_{2s}^2 \beta_{2s}^2/8\pi n_2
u_{2s}$. So strictly speaking, $E_P/E_K =\sigma
(1+\beta_{1s}^2)/2\beta_{1s}^2$.  
In the high-$\sigma$ regime, we have $\beta_{1s} \sim 1$ and the
factor $(1+\beta_{1s}^2)/2\beta_{1s}^2 \simeq 1$. In the low 
$\sigma$ regime, $\beta_{1s} < 1$, but the factor 
$[1+\sigma(1+\beta_{1s}^2)/2\beta_{1s}^2]$ is in any case $\sim 1$. We
therefore neglect the $(1+\beta_{1s}^2)/2\beta_{1s}^2$ factor in the
following discussion.}, so that $E=E_K(1+\sigma)$ or $E_K 
= E/(1+\sigma)$ (see also Zhang \& \Mesz~2002). We follow the
traditional convention to define the Sedov length $l \sim (E/n_1 m_p
c^2)^{1/3}$, where the total energy is adopted. The shell baryon number
density $n_4$ is however defined by $E_K$. The density ratio is 
$n_4/n_1 \sim l^3/[\gamma_4^2 \Delta R^2 (1+\sigma)]$, 
where $\Delta={\rm max}(\Delta_0,R/\gamma_4^2)$ is the thickness of 
the shell, $\Delta_0=c T$ is the initial width of the shell, and $R$ 
is the fireball radius. This holds for both a non-spreading shell 
(where $\Delta= \Delta_0$ is a constant) and for a spreading shell 
(where $\Delta \sim R/\gamma^2$). 

Below we revisit the four critical radii related to the reverse shock
deceleration (Sari \& Piran 1995). In our following discussion, we
assume that a reverse shock exists. The asymptotic behaviors at
$\sigma \gg 1$ for various correction factors (presented in
parentheses) therefore are valid for the $\sigma$ range in which
the reverse shock forming condition is satisfied. 

\begin{figure}
\epsscale{.80}
\plotone{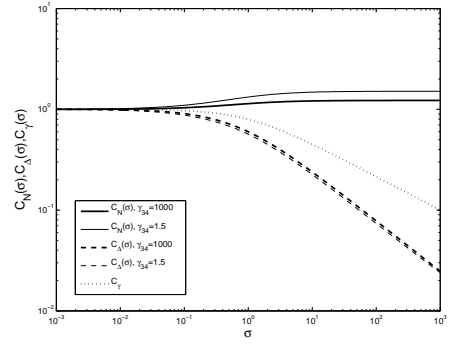}
\caption{
The functions $C_N(\sigma)$, $C_\Delta(\sigma)$ and $C_\gamma$
calculated for both $\gamma_{34}=1.5$ and $\gamma_{34}=1000$. 
\label{fig:Cs}}
\end{figure}

1. The fireball radius for the relative Lorentz factor between the
reverse shock upstream and downstream (i.e. $\gamma_{34}$) to
transform from the Newtonian regime to the relativistic regime can be 
estimated according to $\gamma_4^2 \sim (n_4/n_1) F$, which gives 
\be
R_N \sim \frac{l^{3/2}}{\Delta^{1/2} \gamma_4^2} C_N~,
\ee
where
\ba
C_N(\gamma_{34},\sigma) \simeq C_N(\sigma) =
\left[\frac{F(\sigma)}{1+\sigma} \right]^{1/2} \sim 1, & (\propto
\sigma^0) 
\ea
is the correction factor of $R_N$ with respect to the $\sigma=0$ case.
Since both $F(\sigma)$ and $(1+\sigma)$ have the same asymptotic
behavior ($\propto \sigma$) at high $\sigma$, the final correction
factor $C_N$ is always of order unity throughout (see
Fig.\ref{fig:Cs}), and we will neglect it in the following
discussions. 

2. The radius where the reverse shock crosses the shell is approximately 
(Sari \& Piran 1995) $R_\Delta = [\Delta/(\beta_4 - \beta_2)] 
[1-(\gamma_4/\gamma_3) (n_4/n_3)]$. In the $\sigma=0$ case, the factor 
$[1-(\gamma_4/\gamma_3) (n_4/n_3)]$, which delineates the relative
compression factor due to shock crossing\footnote{If one assumes that
after shock crossing a shell with width $\Delta$ becomes $\Delta'$,
this parameter is simply $(\Delta-\Delta') / \Delta$.}, is a factor
ranging from 
$1/2-6/7$, which was neglected for order-of-magnitude estimates (Sari
\& Piran 1995). For arbitrary $\sigma$ values, this parameter is
$\sigma$-sensitive (through the dependence of $f_b(\sigma)$, see
eq.[\ref{fb}]), i.e., becomes $\ll 1$ when $\sigma \gg 1$, so we 
can not drop it out. Following the similar procedure as in Sari \&
Piran (1995), and replacing $n_4/n_1$ (in the $\sigma=0$ case) by
$(n_4/n_1)F$, one finally has 
\be 
R_\Delta \sim \gamma_4 \Delta \left[\frac{n_4}{n_1} F(\sigma)
\right]^{1/2} \left(1-\frac{\gamma_4
n_4}{\gamma_3 n_3}\right) \sim \Delta^{1/4} l^{3/4} C_\Delta~,
\label{RDelta} 
\ee 
where
\ba
C_\Delta(\gamma_{34},\sigma) & \simeq & C_\Delta(\sigma) =
\left[\frac{F(\sigma)}{1+\sigma}\right]^{1/4} \left[1-\frac{\gamma_4 
n_4}{\gamma_3 n_3} \right]^{1/2} \nonumber \\
& \sim & \left( 1-\frac{\gamma_4 
n_4}{\gamma_3 n_3} \right)^{1/2}, ~~~~~~~ (\propto \sigma^{-1/2})~
\label{CDelta}
\ea
is the correction factor of $R_\Delta$ with respect to the $\sigma=0$
case. This correction factor suggests that the reverse shock crosses
the shell faster when $\sigma$ becomes larger.

3. The conventional ``deceleration radius'' (for the thin shell case)
is modified in the 
high-$\sigma$ regime. According to eq.(\ref{EB}), the ratio of the
comoving magnetic fields in region 4 and 3 is
$B_4/B_3=u_{3s}/u_{4s}$. The lab-frame ratio of the Poynting flux
energy in both regions can be written as
\be
\frac{E_{P,4}}{E_{P,3}}=\frac{\gamma_4 (B_4^2/n_4)}{\gamma_3 (B_3^2
/n_3)} =\frac{u_{3s}\gamma_4}{u_{4s}\gamma_3} \sim 1~,
\label{Econservation}
\ee 
where $\sim$ applies in the $\sigma \gg 1$ limit so that $u_{4s} \sim
\gamma_{4s}$ and $u_{3s} \sim \gamma_{3s}$. Since
$\gamma_{4s}/\gamma_{3s} = \gamma_4/\gamma_3$ (both have the same
relation with $\gamma_{34}$ (eq.[\ref{gam34}]),
eq.(\ref{Econservation}) is naturally derived. We have calculated this
ratio numerically and found that in the high-$\sigma$ limit, the
difference between this ratio and unity is a small quantity comparing
$\sigma^{-1}$. This manifests that shocks in the 
high-$\sigma$ limit only effectively dissipate the kinetic energy (in
the baryonic component) in the
upstream, and the Poynting energy (in the lab frame) essentially
remains the same. As a result, one should define the deceleration
radius (for the thin shell case) using $E_K$ alone, so that
\be 
R_\gamma \sim \frac{l}{\gamma_4^{2/3}(1+\sigma)^{1/3}}
\sim \frac{l}{\gamma_4^{2/3}} C_\gamma~, 
\ee 
where 
\ba
C_\gamma(\sigma) = (1+\sigma)^{-1/3} & (\propto \sigma^{-1/3})~.
\ea
Notice that the radius $R_\gamma$ is still the radius where the
fireball collects $1/\gamma_4$ of fireball rest mass\footnote{This
could be derived using energy conservation equation in the lab frame
before and after the shock crossing(s), i.e. $\gamma_4 (M_0c^2 +U_{\rm
B,0}) +M_{\rm ISM}c^2=\gamma_2 (M_0c^2+\gamma_2 M_{\rm ISM}c^2
+U_{\rm B})$, where $M_0$ is the mass in the original
ejecta, $M_{\rm ISM}$ is the collected ISM mass as the reverse shock
crosses the shell, $U_{\rm B,0}$ is the initial comoving magnetic
energy, and $U_{\rm B}$ is the comoving magnetic energy after shock
crossing. This gives $(\gamma_4-\gamma_2)M_0c^2 + (E_{P,4}-E_{P,3})
=(\gamma_2^2-1) M_{\rm ISM}c^2 $, where $E_{P,4}=\gamma_4 U_{\rm B,0}$
and $E_{P,3}=\gamma_2 U_{\rm B}$. According to equation
(\ref{Econservation}), the term $(E_{P,4}-E_{P,3})$ drops out from the
energy conservation equation, so that the equation is effectively the
familiar hydrodynamical one with the total energy being
$E_K=E/(1+\sigma)$.}. 

In the above discussion, we already assumed that a reverse shock
exists. In order to satisfy the condition (\ref{RS-condition-1}) at
$R_\gamma$, one can give a more explicit constraint on $\sigma$, which
reads
\be
\sigma < 100 \left(\frac{\gamma_4}{300}\right)^4 \left(\frac{T}{10
~{\rm s}}\right)^{3/2} \left(\frac{E}{10^{52} ~{\rm
ergs}}\right)^{-1/2}~.
\label{RS-condition-2}
\ee
We can see for typical GRB parameters, a reverse shock exists when
$\sigma$ is smaller than several tens to several hundreds.

It is worth noticing that at the deceleration radius, the Poynting
energy is not transferred to the ISM yet. At the end shock crossing,
the magnetic pressure behind the contact discontinuity balances the
thermal pressure in the forward shock crossing. It is not until the
reverse shock disappears during the deceleration phase when the bulk
of magnetic energy in the ejecta 
transfers to the forward shock region. During the
deceleration, the magnetic fields push the contact discontinuity from
behind and transfer energy through $p dV$ work.  Eventually, the
total energy will be transferred to the ISM, so that the late time
afterglow level is still defined by the total energy of the fireball.
The detailed energy transfer process is complicated and will be
studied carefully in a future work (see Zhang \& Kobayashi 2005 for a
brief discussion). This point is relevant
to the calculations of the forward shock emission level, and it will
be further discussed in \S\ref{sec:lc}.

In Fig.\ref{fig:Cs} we numerically plot the functions 
$C_N(\sigma)$, $C_\Delta(\sigma)$ and $C_\gamma(\sigma)$ for both the  
mildly relativistic case ($\gamma_{34} = 1.5$) and the extremely 
relativistic case ($\gamma_{34} = 1000$). We can see that $C_N$ is
insensitive to both $\gamma_{34}$ and $\sigma$, and we will treat it
as a constant of order unity. The correction factor $C_\Delta$
is rather insensitive to $\gamma_{34}$ and is essentially a function
of $\sigma$ only. In the $\sigma \gg 1$ regime, we have $C_\Delta
\propto \sigma^{-1/2}$. By definition, $C_\gamma$ is the function of
$\sigma$ only.

4. Finally, the radius where the shell spreads is still defined by
\be
R_s \sim \gamma_4^2 \Delta_0~.
\ee

Taking the convention to define (Sari \& Piran 1995)
\be
\xi \equiv (l/\Delta)^{1/2} / \gamma_4^{4/3}~,
\label{xi}
\ee 
one has the following equation 
\be
\frac{R_N}{\xi}
= \frac{R_{\gamma}}{C_\gamma} = \frac{\xi^{1/2}}{C_\Delta} R_\Delta =
\xi_0^2 R_s~, 
\label{allRs}
\ee
where 
\be
\xi_0 \equiv \frac{(l/\Delta_0)^{1/2}}{\gamma_4^{4/3}} =\left(
\frac{t_\gamma}{T}\right)^{1/2}
\label{xi0}
\ee 
is the $\xi$ value for $R \leq R_s$ (i.e. no
spreading occurs), where 
\be
t_\gamma \equiv \frac{R_\gamma}{C_\gamma \gamma_4^2 c} \sim
\frac{l}{\gamma_4^{8/3} c}.
\label{tgam}
\ee
Notice that this notation is exactly the same as in the $\sigma=0$
case, and the total energy $E$ is used (in defining $l$). This makes
$t_\gamma$ a constant not depending on $\sigma$, which is convenient
for the discussions of the various parameter regions in the next
section. 

\subsection{Parameter regions: thick vs. thin shell regimes
\label{sec:regimes}} 

\begin{figure}
\epsscale{.80}
\plotone{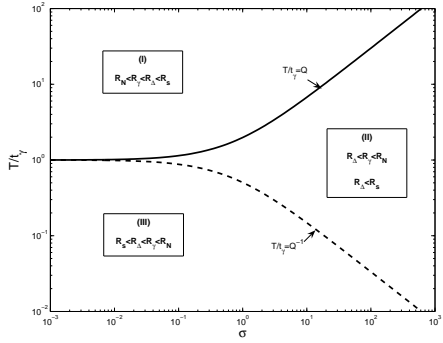}
\caption{
Parameter regimes in the $(T/t_\gamma)-\sigma$ space. 
\label{fig:lines}}
\end{figure}

Equating the four critical radii defines six critical lines in the
$\xi_0-\sigma$ or the $T/t_\gamma - \sigma$ space. This is justified
by the fact that the spreading regime (which makes $\xi$ deviating
from $\xi_0=(t_\gamma/T)^{1/2}$) always happens below the critical
lines, and hence, does not influence the location of the critical
lines. The six critical lines are 
\ba 
R_\gamma \sim R_\Delta, & \frac{T}{t_\gamma} \sim \left(
\frac{C_\gamma}{C_\Delta}\right)^{4} \sim Q, &
(\propto \sigma^{2/3})~; \nonumber \\ 
R_N \sim R_\Delta, & \frac{T}{t_\gamma} \sim 
{C_\Delta}^{-4/3} \sim Q, & (\propto\sigma^{2/3})~; \nonumber \\
R_N\sim R_\gamma, & \frac{T}{t_\gamma}\sim C_\gamma^{-2} \sim
Q, & (\propto\sigma^{2/3})~; 
\nonumber \\ 
R_N \sim R_s, & \frac{T}{t_\gamma}\sim 1, &
(\propto\sigma^0)~; \nonumber \\ 
R_\gamma\sim R_s, & \frac{T}{t_\gamma}\sim C_\gamma \sim Q^{-1/2}, &
(\propto\sigma^{-1/3})~; 
\nonumber \\  
R_\Delta\sim R_s, & \frac{T}{t_\gamma}\sim C_\Delta^{4/3} \sim Q^{-1},
& (\propto\sigma^{-2/3})~.
\label{lines}
\ea
We can see that the first three lines have a same asymptotic behavior
at high-$\sigma$, and calculations show that they essentially coincide
with each other. Hereafter we define
\ba
Q(\sigma) \equiv [C_\gamma(\sigma)]^{-2} \sim C_\Delta^{-4/3} \sim
(C_\gamma/C_\Delta)^4~ & (\propto \sigma^{2/3}), 
\label{Q}
\ea
so that $C_\Delta \sim Q^{-3/4}$ and $C_\gamma \sim Q^{-1/2}$,
and eq.(\ref{allRs}) can be re-written as 
\be
\frac{R_N}{\xi}
= Q^{1/2}R_{\gamma} = \xi^{1/2}Q^{3/4} R_\Delta =
\xi_0^2 R_s~, 
\label{allRs2}
\ee
In principle, changing the order between $R_N$ and $R_s$ and between
$R_\gamma$ and $R_s$ does not lead to essential modifications of the
shock crossing and deceleration physics, so that the 4th and 5th lines
in eq.(\ref{lines}) are not crucial. We therefore have two essentially
lines that separate three physical regimes in the $T/t_\gamma -
\sigma$ space (Fig.\ref{fig:lines}).

Region (I): the thick shell regime. This is the region where
$T/t_\gamma > Q$ is satisfied. In this region, one has
$R_N<R_\gamma<R_\Delta<R_s$. The downstream becomes relativistic
with respect to the upstream before the reverse shock crosses 
the shell (Sari \& Piran 1995).  The full deceleration occurs at the
end of shock crossing, i.e. at $R \sim R_\Delta$ (Kobayashi et
al. 1999). Given a certain observed central engine activity time
$T$, a total energy $E$, and an ambient density $n$, one can
also define a critical Lorentz factor 
\be
\gamma_c \simeq 125 E_{52}^{1/8} n^{-1/8} T_2^{-3/8} Q^{3/8} 
\left(\frac{1+z}{2}\right)^{3/8}~, 
\label{gammac}
\ee
where some refined coefficients and the cosmological time dilation 
factor are explicitly taken into 
account in order to get the numerical value (here $z$ is the GRB
redshift). For $\gamma_0 > \gamma_c$, 
one is in the thick shell regime, while for $\gamma_0 < \gamma_c$, one 
is in the thin shell regime. 
The function $Q(\sigma)$ is plotted as the top curve in
Fig.\ref{fig:lines}. In the $\sigma \ll 1$ regime, $Q(\sigma) \sim 1$.
For $\sigma \gg 1$, we have $Q(\sigma) \propto \sigma^{2/3}$, and hence,
$\gamma_c \propto \sigma^{1/4}$. We can see that given same values
for the other parameters, the parameter space for the thick shell
regime is greatly reduced when $\sigma$ is high. More bursts are in
the thin shell regime.

Region (II): the non-spreading thin-shell regime. This region is
defined by $Q^{-1} < (T/t_\gamma) < Q$, in which $R_\Delta < R_\gamma
< R_N$ and $R_\Delta < R_s$ are satisfied.
The common features of this regime are that the
reverse shock crosses the shell (at $R_\Delta$) before noticeable
deceleration occurs (at $R_\gamma$, e.g. the Lorentz factor is reduced
by roughly a factor of 2), that the relative speed between
the upstream and the downstream never becomes
relativistic, and that the shell does not spread during the first
shock crossing. The separation between $R_\Delta$ and $R_\gamma$ leads
to some 
novel features for the reverse shock emission. During the first shock
crossing the shell is heated so that electrons start to emit. However,
after the first shock crossing, the shell is not decelerated
significantly. It is difficult to delineate the detailed process at
this stage, but a rough picture is that higher order shocks may form
and bounce back and forth between the inner edge of the shell and the
contact discontinuity. This happens until the shell reaches
$R_\gamma$. It is likely that the shell remains heated by the
multi-crossing of shocks and electrons keep radiating at a high level
for an extended period of time. One then expects a broad reverse shock
emission peak, which is a novel phenomenon in the high-$\sigma$
thin-shell regime. Notice that when $\sigma$ is very large, a reverse
shock may not form at $R_\Delta$ at all. However, since $R_\Delta$ is
the smallest in the problem, whenever a reverse shock forms, it
quickly crosses the shell in a radius of $R_\Delta$, and the above
discussion is still valid.

Region (III): the spreading thin-shell region. This is defined by
$(T/t_\gamma) < Q^{-1}$, in which $R_s < R_\Delta < R_\gamma <
R_N$ is satisfied. Since $R_\Delta < R_\gamma$, again multi-crossing
of shocks are needed to slow down the ejecta, and the downstream never
becomes relativistic with respect to the upstream. The novel feature
in this region compared with 
the region (II) is that the shell starts to spread before shock 
crossing, so that the three radii have the relationship
\be
C_\gamma^{2/3} R_N = R_\gamma = (C_\gamma/C_\Delta)^{2} R_\Delta, 
\label{spreading}
\ee
or 
\be
Q^{-1/3} R_N \simeq R_\gamma \simeq Q^{1/2} R_\Delta~. 
\label{spreading2}
\ee
We can see that the triple coincidence $R_N = R_\gamma = R_\Delta$ in 
the thin shell regime (Sari \& Piran 1995) is only valid when $\sigma$
is small. According to Fig.\ref{fig:lines}, this practically happens
when $\sigma \leq 0.01$.

\subsection{Critical times}

\begin{figure}
\epsscale{.80}
\plotone{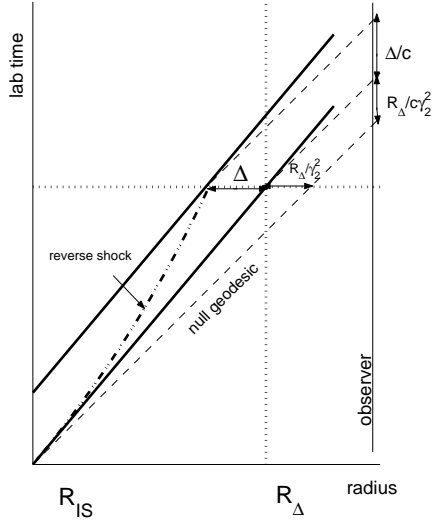}
\caption{
The space-time diagram of the GRB fireball evolution. Solid lines are
the world lines of both ends of the shell, and the dashed lines are
the world lines of light. The lowest dashed line is the first
light (from the internal shock) that reaches the observer, and the
other two dashed lines indicate the light emitted from both ends of
the shell at the shock-crossing radius, $R_\Delta$. The observed
shock crossing time is the sum of the delay time
$R_\Delta/(2\gamma_2^2 c)$ and $\Delta/c$. To order of magnitude
estimate, the compression of the shell after shock crossing is not
taken into account.
\label{fig:times}}
\end{figure}

We finally derive the shock crossing time $t_\times$ and the
deceleration time $t_{dec}$ as measured by the observer. In the
literature to study the $\sigma=0$ regime, 
$t_{dec}=t_\times \sim {R_\Delta}/{\gamma_2^2 c} \sim {\rm
max}(T,t_\gamma C_\gamma) \sim {\rm max}(T,t_\gamma)$ has been
conventionally adopted (noticing $C_\gamma =1$ when $\sigma=0$). 
When an arbitrary $\sigma$ value is adopted, there are further
complications to quantify these critical times. First,
although in the thick shell regime (I) $t_{dec}=t_\times$ is still
valid, in the thin shell regimes (II and III) the deceleration radius
$R_\gamma$ is larger
than the shock crossing radius $R_\Delta$, so that $t_\times <
t_{dec}$. Second, the time scale $R_\Delta/\gamma_2^2 c$ only
describes the delay time scales for the emission coming from the
radius $R_\Delta$ with respect to the emission from the internal shock
radius, for an infinitely thin shell. A more precise description of the
reverse shock emission peak time should include the thickness of the
radiation region. The real shock crossing time should correspond to
the epoch when the emission from the end of the shell reaches the
observer (see Fig.\ref{fig:times} for illustration). This gives
\be
t_\times \sim \frac{R_\Delta}{\gamma_2^2 c}+\frac{\Delta}{c}~.
\label{tx}
\ee
Here $\Delta={\rm max}(\Delta_0,R/\gamma^2)$, so that our definition
is valid throughout the $(T/t_\gamma) - \sigma$ space. 
In the $\sigma \ll 1$ limit, we always have ${R_\Delta}/{\gamma_2^2 c}
\sim \Delta/c$ so that to order of magnitude estimate, one can drop
the latter term. In the $\sigma \gg 1$ limit, however, in certain
regimes one could have ${R_\Delta}/{\gamma_2^2 c} \ll \Delta/c$. The
correction factor introduced here is therefore essential to delineate
the reverse shock behavior in the high-$\sigma$ regime. We notice that
Nakar \& Piran (2004) recently also noticed this correction within 
the context of $\sigma=0$ regime, although their eq.(2) is slightly
different from our definition. 

In the thick shell regime (I) the reverse shock is relativistic at
the crossing radius, and one has $\gamma_2^2 =\gamma_4
[({n_4}/{n_1}) F]^{1/2}$. Using the definition of $R_\Delta$
(eq.[\ref{RDelta}]), one has $R_\Delta/\gamma_2^2c \sim T C_\Delta^2
\sim T Q^{-3/2}$.
Since $\Delta/c=\Delta_0/c=T$, with eq.(\ref{RDelta}), we
have  $t_\times({\rm I}) = t_{dec}({\rm I}) \sim
T(1+Q^{-3/2})$. In the region (II), i.e., the non-spreading thin shell
regime, one has $\gamma_2 \sim \gamma_4$ and $\Delta = \Delta_0$. Using
eqs.(\ref{allRs}) and (\ref{tgam}), we get $t_\times({\rm II}) \sim
t_\gamma^{3/4} 
{T}^{1/4} C_\Delta + T \sim t_\gamma^{3/4} T^{1/4}
Q^{-3/4} +T$. In the region (III), i.e. the spreading thin
shell regime, one has $\gamma_2 \sim \gamma_4$ and $\Delta/c =
R_\Delta/\gamma_4^2 c$. With eq.(\ref{spreading}), one has
$t_\times \sim 2t_\gamma (C_\Delta^2 / C_\gamma)\sim 2t_\gamma
Q^{-1}$. 

We define the deceleration time $t_{dec}$ as the epoch when the
fireball is significantly decelerated. For the thick shell regime (I),
this coincides with the shock crossing time. For the thin shell
regimes (II, III), it is defined when the fireball reaches $R_\gamma$.
Following the same argument to derive (\ref{tx})
(Fig.\ref{fig:times}), one can generally define
\begin{equation}
t_{dec} \sim \left\{
  \begin{array}{@{\,}lc} 
    \frac{R_\Delta}{\gamma_2^2c}+\frac{\Delta}{c} = t_\times, &
\mbox{(I)}  \\ 
\frac{R_\gamma}{\gamma_2^2c}+\frac{\Delta}{c} = t_\gamma C_\gamma
+\frac{\Delta}{c}~, & \mbox{(II, III)}
  \end{array} 
  \right. 
\label{tdec}
\end{equation} 
where we have interchanged $\gamma_2$ and $\gamma_4$ for the thin
shell regimes. For the regime II, one has $t_{dec}({\rm II}) \sim
t_\gamma Q^{-1/2}+T$, while for the regime III, one has $t_{dec}({\rm
III}) \sim t_\gamma (Q^{-1/2}+Q^{-1})$.

For convenience, we collect the critical times as the following,
\begin{equation}
t_\times \sim \left\{
  \begin{array}{@{\,}lc}
    T(1+Q^{-3/2}), & \mbox{(I)}  \\
    t_\gamma^{3/4} {T}^{1/4} Q^{-3/4} +T,
                & \mbox{(II)} \\
    2t_\gamma Q^{-1}, & \mbox{(III)}
  \end{array}
  \right. ~,
\label{txall}
\end{equation}
and
\begin{equation}
t_{dec} \sim \left\{
  \begin{array}{@{\,}lc}
    T(1+Q^{-3/2}), & \mbox{(I)}  \\
    t_\gamma Q^{-1/2} +T, & \mbox{(II)} \\
    t_\gamma(Q^{-1/2}+Q^{-1}), & \mbox{(III)} 
  \end{array}
  \right. ~.
\label{tdecall}
\end{equation}
As demonstrated above, in the thin shell regime, the reverse shock
emission should show a broad peak due to the separation between
$t_\times$ and $t_{dec}$. The width of the peak of the reverse shock
peak is defined as
\begin{equation}
(t_{dec}-t_\times) \sim \left\{
  \begin{array}{@{\,}lc}
    0, & \mbox{(I)}  \\
    t_\gamma [Q^{-1/2}-(T/t_\gamma)^{1/4} Q^{-3/4}], & \mbox{(II)} \\
    t_\gamma(Q^{-1/2}-Q^{-1}), & \mbox{(III)} 
  \end{array}
  \right. ~.
\label{width}
\end{equation}

\section{Synchrotron emission and early afterglow lightcurve}

\subsection{Particle acceleration}

The hydrodynamical solution presented above is generic and lays a
solid foundation for calculating synchrotron emission in the reverse
shock and the early afterglow lightcurves. In this section, we will
turn to the less certain aspects of the problem, i.e. particle
acceleration and synchrotron emission from the reverse shock. In the
conventional $\sigma=0$ models, particles (ions and electrons) are
assumed to be accelerated from the collision-less relativistic shocks
through the first-order and probably also second-order stochastic Fermi
acceleration mechanisms
(Fermi 1949; Blandford \& Eichler 1987). In the standard afterglow
models, accelerated electrons are assumed to have a power law
distribution with $dN_e/d\gamma_e \propto \gamma_e^{-p}$. Numerical
simulations of particle acceleration in relativistic shocks confirm
this simple treatment (e.g. Gallant et al. 1992; Achterberg et
al. 2001) in the $\sigma=0$ limit.
For MHD shocks as discussed in this paper, more physical
processes enter the problem. For example, in the $90^{\rm o}$ shock
problem discussed in this paper, the existence of the electrostatic
potential in the shock front plane and the influence of the Lorentz
force exerted on the particles by the magnetic and electric fields in
the upstream tend to trap ions in the shock plane. The influence of
these effects on shock acceleration is unclear, although some
investigations in this direction have started (e.g. Double et
al. 2004; Spitkovsky \& Arons 2004). Detailed treatments of particle
acceleration in the high-$\sigma$ regime is beyond the scope of the
current paper. Lacking a detailed model, here we simply extend the
approach used in the $\sigma=0$ regime to arbitrary $\sigma$ values, 
i.e. we assume a power law distribution of electron energy and assign
the equipartition parameters $\epsilon_e$ and $\epsilon_B$ for the
electrons and magnetic fields, respectively. Such an approximation is
proven valid when $\sigma$ is small, but may progressively become not
good enough as $\sigma$ increases, especially when $\sigma$ achieves
very large values. The lightcurves we calculate below nonetheless
provide a first-order picture on how the reverse shock emission level
depends on $\sigma$. 

\subsection{Magnetic fields}

We follow our previous approach (ZKM03) to compare the flux
level between the reverse shock peak and the forward shock peak. This
is because, when studying the ratio of the peak flux levels, only the
ratios of the micro-physics parameters (e.g. $g(p)=(p-2)/(p-1)$,
$\epsilon_e$, $\epsilon_B$, etc.) matter, and one does not need to
invoke the absolute values of those parameters which are rather
uncertain. A crucial parameter to study synchrotron spectrum is the
comoving magnetic field strength in both shocked regions. For the
forward shock region, since the medium is usually not magnetized, the
downstream magnetic field is usually quantified by a fudge parameter
$\epsilon_{B,f}$, which reflects the strength of the magnetic field
presumably generated in-situ due to a certain plasma instability
(e.g. Medvedev \& Loeb 1999). This magnetic field is randomly
oriented, as has been supported by the observed weak-polarization
level for the optical afterglow emission (e.g. Covino et al. 2003 for
a review). The strength of this magnetic field component is low, with
$\epsilon_{B,f} \sim (0.01-0.001)$, as inferred from broadband
afterglow fits (Panaitescu \& Kumar 2002; Yost et al. 2003).
For the reverse shock, in the current model the magnetic field is
predominantly due to the compression of the upstream magnetic field.
Its level depends on the $\sigma$ value of the upstream, and the field
is globally structured, so that the optical flash due to the reverse
shock emission (such as the ones observed from GRB 990123 and GRB
021211) should have been strongly polarized (see also Granot \&
K\"onigl 2003; Fan et al. 2004a; Sagiv, Waxman \& Loeb 2004).

The forward shock comoving magnetic energy density is defined by
\be
\frac{B_f^2}{8\pi} = \frac{B_2^2}{8\pi} = e_2 \epsilon_{B,f}~,
\ee
where $\epsilon_{B,f}$ is the conventional magnetic equipartition
parameter in the afterglow theory which delineates the fraction of the 
total internal energy that is distributed to magnetic energy.
In the reverse shock region, the comoving magnetic energy density is
dominated by the shock-compressed upstream magnetic field, and can be
denoted as
\be
\frac{B_r^2}{8\pi} = \frac{B_3^2}{8\pi} =(f_c-1) \frac{e_3}{3}~, 
\label{Br1}
\ee
where $f_c\equiv 1+p_{b,3}/p_3$ (eq.[\ref{fc}]) has been used.
For easy comparison (with respect to the conventional definition of
$\epsilon_{B,f}$), we can write (\ref{Br1}) as
\be
\frac{B_r^2}{8\pi} = e_2 \bar \epsilon_{B,r}~,
\label{Br2}
\ee
where
\be
\bar \epsilon_{B,r} \equiv \frac{(f_c-1)}{3 f_c}
\label{epsilonBr}
\ee
is an artificial parameter to simplify the discussions.
With this definition, we can write
\be
{\cal R}_B\equiv \frac{B_r}{B_f} =
\left(\frac{\bar\epsilon_{B,r}}{\epsilon_{B,f}} 
\right)^{1/2}~.
\ee
This is an important parameter which delineates the ratio of the
magnetic field strength in the reverse shock and forward shock
regions. This ratio has been found to be larger than unity in GRB
990123 and GRB 0201211 (e.g. ZKM03), and discussion of this parameter
is very essential to quantify the relative emission properties of both
shocks. 

\begin{figure}
\epsscale{.80}
\plotone{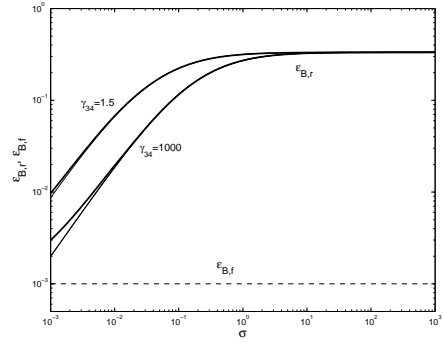}
\caption{
The equivalent ``magnetic equipartition parameter'' in the reverse
shock region, $\bar\epsilon_{B,r}$, as a function of $\sigma$,
calculated for both $\gamma_{34}=1000$ and $\gamma_{34}=1.5$. The
dashed line is the assumed magnetic equipartition parameter in the
forward shock region, i.e., $\epsilon_{B,f} \sim 0.001$, which is also 
the ``bottom level'' for the reverse shock magnetic field in the
low-$\sigma$ regime. The thin solid lines are calculated completely
from eq.(\ref{epsilonBr}), while the thick solid lines include the
contribution of the random field in the low-$\sigma$ regime. The fact
that the parameter $\bar\epsilon_{B,r}$ approaches 1/3 in the
high-$\sigma$ regime is only an artificial effect resulting from the
definition of $\bar\epsilon_{B,r}$ (eq.[\ref{epsilonBr}]).
\label{fig:epsB}}
\end{figure}

In Fig.\ref{fig:epsB}, we plot $\bar\epsilon_{B,r}$ as a function of
$\sigma$ for both $\gamma_{34}=1000$ and $\gamma_{34}=1.5$. We can see 
that it increases with $\sigma$ initially, and saturates at a value
1/3 as $\sigma \gg 1$. The asymptotic behavior is already obvious in
eq.(\ref{epsilonBr}). We note that the number 1/3 is a pure artificial
effect given the definition of $\bar \epsilon_{B,r}$ (eq.[\ref{Br2}]).
The ``real'' magnetic equipartition factor in the region 3 approaches
unity when $\sigma \gg 1$. In the figure we also plotted the
magnetic equipartition parameter in the forward shock region, i.e.,
$\epsilon_{B,f} \sim 0.001$. This level could also be regarded as the
``bottom level'' in the $\sigma \ll 1$ regime for the reverse shock
region. The thick lines in
Fig.\ref{fig:epsB} are the ``total'' $\epsilon_B$ in the reverse shock
region (which include both the amplified structured field component
and the random field component), which saturates to $\epsilon_{B,f}$ in
the low-$\sigma$ regime.

An interesting conclusion drawn from Fig.\ref{fig:epsB} is that 
no matter what $\sigma$ or $\gamma_{34}$ values are taken, the ratio
$\bar\epsilon_{B} / \epsilon_{B,f}$ is at most $\sim 300$ for
$\epsilon_{B,f} \sim 0.001$. Or effectively, the reverse-to-forward
shock magnetic field ratio ${\cal R}_B$ can not be significantly
larger than 15. Since this value was inferred from the case of GRB
990123 (ZKM03), we tentatively conclude that the ejecta in GRB 990123
has $\sigma > 0.1$ where $\bar\epsilon_{B,r}$ reaches its peak value. 
The absolute values of $\epsilon_{B,r}$ and $\epsilon_{B,f}$ are
consistent with those obtained from the detailed modeling. For example,
$\epsilon_{B,f} \sim 7.4\times 10^{-4}$ was inferred by Panaitescu 
\& Kumar (2002), while $\epsilon_{B,r} / \epsilon_{B,f} \sim
15^2 \sim 225$ was inferred by ZKM03, so that $\epsilon_{B,r} \sim
0.17$. This is close to the maximum $\bar\epsilon_{B,r}$ we have
calculated. 

\subsection{Lightcurve peak times and peak fluxes}

Before describing the detailed process of calculating afterglow
lightcurves, it is informative to define the so-called peak times and
peak fluxes. An early optical lightcurve usually consists two peaks
(ZKM03 and references there in), i.e., a forward shock peak which
corresponds to the epoch when the typical synchrotron frequency
crosses the band (Sari et al. 1998; Kobayashi \& Zhang 2003a), and a
reverse shock peak at which the flux achieves the maximum and starts
to decay thereafter. This corresponds to the time when no more shock
heating is available and the shell starts to cool adiabatically.
The condition for a reverse shock to exist is expressed in equations
(\ref{RS-condition-1}) and (\ref{RS-condition-2}), and in this paper
we focus on the situation when such a condition is satisfied.
We will denote the reverse (forward) shock peak times
and peak fluxes as $t_{p,r}$ ($t_{p,f}$) and $F_{\nu,p,r}$
($F_{\nu,p,f}$), respectively. 

For the $\sigma=0$ case, the first shock crossing time and the shell
deceleration time coincide, so that $t_{p,r}=t_\times=t_{dec}$.
This is still the case when $\sigma$ is larger as long as the shell is
in the thick shell regime (I). For thin shell cases (II and III),
however, this is no longer the case, and $t_\times$ and $t_{dec}$
separate from each other (eqs.[\ref{tdec}],[\ref{tdecall}]). For easy
discussion, hereafter we 
define the time $t_{dec}$ as the reverse shock peak time, and its
corresponding afterglow flux as reverse shock peak flux, i.e. 
\ba
t_{p,r} & = & t_{dec} \nonumber \\
F_{\nu,p,r} & = & F_{\nu}(t_{dec})
\ea
For $t>t_{p,r}$, the shell cools adiabatically, and a decaying
lightcurve results\footnote{When $\sigma$ is very large, additional
heating for the shell may still happen if the post-shock-crossing
energy transfer process time scale is short enough.}. The $t<t_{p,r}$
case is a little more 
complicated. For the thick shell case (I), since $t_\times$ coincides
with $t_{p,r}$, it is a rising lightcurve due to shock crossing. For
the thin shell cases (II and III), as demonstrated above, the shell is
heated at first shock crossing and remains heated
until it is significantly decelerated. This results in a broad reverse 
shock peak which starts at $t_\times$ and ends at $t_{p,r}$. Much more
detailed studies are needed to reveal the physics during this
stage, but to a first-order estimate, in this paper we assume that the
heating level between $t_\times$ and $t_{p,r}$ is roughly the same, so 
that the lightcurve shows a plateau during this period. 

To calculate the synchrotron radiation flux, one needs to quantify the
comoving random Lorentz factor of the leptons. We still take the 
convention to assume that the shock accelerated leptons have a single
power-law distribution with the indices $p_f$ and $p_r$ for the
forward and the reverse shocks, respectively, and that they occupy a
fraction $\epsilon_{e,f}$ and $\epsilon_{e,r}$ of the total thermal
energy in the forward shock region (which is $e_2$) and in the
reverse shock region (which is $e_3$), respectively. For the reverse
shock region, the lepton density may be enriched by the presence of
pairs generated in the prompt emission phase (e.g. Li et al. 2003b).
The pair-multiplicity parameter 
\be
y \equiv {(N_{b}+N_\pm)}/{N_{b}} \geq 1
\label{y}
\ee 
may be of order unity or mildly large in the low-$\sigma$
regime (depending on the compactness of the region when the prompt
gamma-rays are emitted, e.g. Kobayashi, Ryde \& MacFadyen 2002; \Mesz~
et al. 2002), and could be very large in the high-$\sigma$ regime
(e.g. Zhang \& \Mesz~2002).
In this paper, we are mainly focusing on the novel features introduced
by the $\sigma$ parameter, and will take $y\sim 1$ in the following
calculations. The $y$-dependences are included in the expressions and
their implication will be discussed in \S\ref{sec:lc}.

The minimum comoving electron energy in the region ``$i$'' (2 or 3) is
$\gamma_{e,m,i}=(\epsilon_{e,i}/y_i) (e_i/n_i m_p c^2)$ $g(p_i) (m_p/m_e)$,
where $g(p)=(p-2)/(p-1)$ (assuming $p>2$). Taking the values at the
first shock crossing time $t_\times$, one gets
\be
\frac{\gamma_{e,m,r}(t_\times)}{\gamma_{e,m,f}(t_\times)}=
\left(\frac{\epsilon_{e,r}}{y\epsilon_{e,f}} \frac{g_r}{g_f}\right) f_a 
\frac{\gamma_{34}(t_\times)-1}{\gamma_{2}(t_\times)-1} \sim {\cal R}_e f_a
\frac{\gamma_0}{\gamma^2_\times} y^{-1}~,
\label{Rgamx}
\ee
where we have defined 
\be
{\cal R}_e \equiv \frac{\epsilon_{e,r} g_r}{\epsilon_{e,f} g_f}~, 
\ee
used $\gamma_{34} \sim \gamma_4 / \gamma_2(\times)$ (which is valid
for both thick and thin shells), and replaced
$\gamma_4$ and $\gamma_2(\times)$ by $\gamma_0$ (which 
means the initial Lorentz factor) and $\gamma_\times$ (the fireball
Lorentz factor at the shock crossing time),
respectively\footnote{Strictly speaking, the last factor
$\gamma_0/\gamma_\times^2$ in eq.(\ref{Rgamx})
should be $(\gamma_0-\gamma_\times)/\gamma_\times^2$. The current
approximation is valid as long as the reverse shock is mildly
relativistic, say, $\gamma_{34} > 1.5$. For an even smaller
$\gamma_{34}$ (which could be possible when $\sigma$ is large enough), 
there should be an additional correction factor (less than unity) in
both eqs.(\ref{Rgamx}) and (\ref{Rgamdec}).}.
This allows the same notation system as in our previous work
(ZKM03). 

We are more interested in the behavior at the reverse shock peak time
(i.e. the deceleration time),
$t_{p,r}$. For the thick shell case, this is simply $t_\times$.
For the thin shell case, after the first shock crossing, the shell is
kept heated by multi-crossing of successive shocks. To first order, we
can take the approximation that the heating 
level in the ejecta during the time period between $t_\times$ and
$t_{p,r}$ is approximately constant, i.e., $e_3/n_3 \propto
t^{0}$. During the same period the random Lorentz factor in the
shocked ISM region also remains constant (since $\gamma_2 \sim
\gamma_4$ before deceleration), we therefore also have
\be
\frac{\gamma_{e,m,r}(t_{p,r})}{\gamma_{e,m,f}(t_{p,r})}
\sim \frac{\gamma_{e,m,r}(t_\times)}{\gamma_{e,m,f}(t_\times)} \sim
{\cal R}_e f_a \frac{\gamma_0}{\gamma^2_\times} y^{-1}~,
\label{Rgamdec}
\ee
which is valid for both the thick and thin shell cases.

Let us denote the total electron numbers in the forward and reverse
shocked region as $N_{e,f}$ and $N_{e,r}$, respectively.  
In the reverse shock region, the total lepton 
number (now includes pairs) is $N_{e,r}=N_{b}+N_\pm = yN_{b}$,
where the definition in eq.(\ref{y}) is used. 
According to eq.(\ref{Econservation}), at the deceleration radius, the
total energy in the forward shock region is defined by $E_K$ alone.
Although the bulk of the Poynting energy is expected to be transferred
to the ISM eventually, shortly after the shock crossing and near the
forward shock peak, this correction may not be significant. Below we
will ignore this process in our calculations of the forward shock
emission, but keeping in mind that the real forward shock emission
level would increase with time, and may be much higher than our
predicted level at later times. When we focus on early afterglow
lightcurves, our calculations should be close to the real emission
level (see more discussions in Zhang \& Kobayashi 2005). A more
careful treatment will be presented in a future work. 

In our approximated treatment, one can write
$E_K =\gamma_0 c^2 (N_{b}m_p+N_\pm m_e) \sim
\gamma_0 N_{b}m_p c^2  \sim N_{e,f}m_p c^2
[\gamma(t_{dec})]^2 \sim N_{e,f}m_p c^2 \gamma_\times^2$,
where we have assumed $y \ll m_p/m_e$, so that the total pair
mass $N_\pm m_e$ is much smaller than the total baryon mass $N_{b}
m_p$. This gives  
\be
\frac{N_{e,r}(t_{p,r})}{N_{e,f}(t_{p,r})} \sim 
y \frac {\gamma_\times^2}{\gamma_0}~.
\label{RNdec}
\ee

The characteristic synchrotron emission frequency is $\nu_m 
\propto \gamma B \gamma_{e,m}^2$, the cooling frequency is $\nu_c \propto
\gamma^{-1} B^{-3} t^{-2}$, and the peak specific flux is $F_{\nu,m}
\propto \gamma B N_e$, where $\gamma$ is the bulk Lorentz factor. 
For the thin shell case, we also make another approximation that $B_3$
keeps constant from $t_\times$ to $t_{dec}$ (so that ${\cal
R}_B(t_\times) \simeq {\cal R}_B(t_{dec})={\cal R}_B$). 
Similar to Kobayashi \& Zhang (2003a) and ZKM03, we can
finally derive the following relations at $t_{dec}$.
\ba
\frac{\nu_{m,r}(t_{p,r})}{\nu_{m,f}(t_{p,r})} & \sim &
\hat\gamma^{-2} {\cal R}_e^2
{\cal R}_B f_a^2 y^{-2}~, \label{Rnum} \\
\frac{\nu_{c,r}(t_{p,r})}{\nu_{c,f}(t_{p,r})} & \sim & 
{\cal R}_B^{-3}~,  \label{Rnuc} \\
\frac{F_{\nu,m,r}(t_{p,r})}{F_{\nu,m,f}(t_{p,r})} & \sim &
\hat\gamma {\cal R}_B y~,
\label{RFnum}
\ea
where
\be
\hat\gamma \equiv \frac{\gamma_\times^2}{\gamma_0}={\rm
min}\left(\gamma_0,\frac{\gamma_c^2}{\gamma_0}\right) \leq 
\gamma_c~.
\label{hatgam}
\ee

Although there are in principle many cases of the reverse shock
emission lightcurves (Kobayashi 2000), within the reasonable
parameter regime the lightcurve behavior only has two variations
depending on whether ${\cal R}_\nu > 1$ or ${\cal R}_\nu < 1$ (ZKM03) 
where
\be
{\cal R}_\nu \equiv \frac{\nu_R}{\nu_{m,r}(t_{p,r})}~.
\label{Rnu}
\ee
In both cases, the ratio between the two peak-time fluxes
\be
{\cal R}_F \equiv \frac{F_{\nu,p,r}}{F_{\nu,p,f}}
\label{RF}
\ee
and the ratio between the two peak times
\be
{\cal R}_t \equiv \frac{t_{p,f}}{t_{p,r}}
\label{Rt}
\ee
can be expressed in terms of $\hat\gamma$, ${\cal R}_B$ and ${\cal
R}_\nu$, respectively, for the $\sigma=0$ case (ZKM03). Here the
forward shock peak time $t_{p,f}$ corresponds to the epoch when 
$\nu_{m,f}$ crosses the band\footnote{When $\sigma$ is large enough,
additional correction factor to $t_{p,f}$ is needed, but this factor
is small if the energy injection time scale is long enough.}.

Below we repeat this process, but focus more on corrections
introduced by the $\sigma$ factor. To further simplify the problem, we 
first estimate the numerical value of ${\cal R}_\nu$. Due to the
complication introduced by the $\sigma$ parameter, one can not coast
${\cal R}_\nu$ into a simple expression as in the $\sigma=0$ case
(e.g. eq.[24] in ZKM03). In any case, using eq.(\ref{Rnum}) and 
the standard expression for $\nu_{m,f}(t)$ (e.g. eq.[1] in Kobayashi
\& Zhang 2003a), one can write
\ba
{\cal R}_\nu & \sim & 800 {\cal R}_B^{-1} f_a^{-2} y^2
{\cal R}_{e}^{-2} \left(\frac{k E_K}{10^{52}~{\rm erg}}\right)^{-1/2}
\left(\frac{\hat\gamma_2}{100}\right)^2 \nonumber \\
& \times &
\left(\frac{\epsilon_{B,f}}{0.001}\right)^{-1/2} 
\left(\frac{\epsilon_{e,f}}{0.1}\right)^{-2} 
\left(\frac{g_f} {1/3}\right)^{-2} \nonumber \\
&\times & \left(\frac{t_{dec}}
{100~{\rm s}}\right)^{3/2} \left(\frac{1+z}{2}\right)^{-1/2}~.
\label{Rnu}
\ea
For $\hat\gamma \leq \gamma_c \sim 125$ (eq.[\ref{gammac}]), ${\cal
R}_e \sim 1$, $y \geq 1$, and $f_a < 1$, the above equation therefore
essentially 
always gives ${\cal R}_\nu > 1$. In the following discussions, we will
not discuss the ${\cal R}_\nu < 1$ case any further (which was also
discussed in ZKM03).

The reverse shock emission lightcurve in the ${\cal R}_\nu > 1$ case is
simple. The lightcurve initially rises and reaches the peak at
$t_\times$. The flux level then keeps essentially constant until 
$t_{p,r}$ (for the thick shell case, both time scales coincident, so
that there is no broadened peak), and starts to decay after
$t_{p,r}$. The temporal indices of each segment of the lightcurve is
also well-defined. In the rising part of the lightcurves, since all
the correction factors introduced by the $\sigma$ parameter are
essentially time-independent, the corrections essentially do not
introduce extra time-dependence on the typical frequencies and the
peak flux of the synchrotron radiation in the reverse shock. The
rising lightcurves essentially
remain unchanged as the $\sigma=0$ case, as has been derived by
Kobayashi (2000). This gives a $\sim 1/2$ temporal index for the thick 
shell case, and a $\sim 5$ temporal index for the thin shell
case\footnote{Detailed numerical calculations result in non-power law
behavior in the rising lightcurve (Fan et al. 2004a).}.
After the deceleration time, the shell cools. The optical 
band is typically in the regime of $\nu_{m,r}(t_\times) < \nu_R <
\nu_{c,r}(t_\times)$. After the deceleration time, one has $\nu_{m,r}
\propto t^{-3/2}$, $F_{\nu,m,r} \propto t^{-1}$ (Kobayashi
2000)\footnote{A more detailed discussion such as that presented in
Kobayashi \& Sari (2000) leads to the similar conclusion.}. Thus
the temporal decay index (i.e. $F_\nu \propto t^{-\alpha}$) is
\be
\alpha=\frac{3p_r+1}{4} \sim 2~,
\ee
where $p_r$ is the electron power-law index in the reverse shock region.

For $t>t_{p,r}$, one has $\nu_{m,f} \propto t^{-3/2}$,
$F_{\nu,m,f} \propto t^0$ (M\'esz\'aros \& Rees 1997a)\footnote{Notice
again that here we have assumed that the energy transfer time scale
from a Poynting flux to the kinetic energy of the ISM long enough. 
This forward shock emission level should be regarded as a lower limit
when the energy transfer process is taken into account.} and $\nu_{m,r}
\propto t^{-3/2}$, $F_{\nu,m,r} \propto t^{-1}$ (Kobayashi
2000). Using the definitions of ${\cal R}_\nu$, ${\cal R}_F$ and 
${\cal R}_t$ (eqs.[\ref{Rnu}-\ref{Rt}]) as well as eqs. (\ref{Rnum})
and (\ref{RFnum}), one can derive\footnote{In ZKM03, we have
defined ${\cal R}_F$ and ${\cal R}_t$ at $t_\times$, but in $\sigma=0$ 
case one has $t_\times=t_{dec}$. For the case of an
arbitrary $\sigma$, the deceleration time $t_{dec}$ is more
fundamental to define the problem.}
\ba
{\cal R}_t & = & \hat\gamma^{4/3} {\cal R}_B^{-2/3} {\cal
R}_\nu^{-2/3} ({\cal R}_e^{-4/3}y^{4/3} f_a^{-4/3})~, \label{Rt2} \\
{\cal R}_F & = & \hat\gamma {\cal R}_B {\cal R}_\nu^{-2(\alpha-1)/3} (y)
~ \label{RF2}.
\ea
These are valid for all the three parameter regions in
Fig.\ref{fig:lines}. Comparing with eqs.(12,13) in ZKM03,
the extra correction factors are presented in
parenthesis. Notice that the correction factors ${\cal R}_e$ and $y$
should also exist in $\sigma=0$ case, but we have previously assumed
them to be unity. The extra $\sigma$-dependent correction factors are
$f_a^{-4/3}$ and ${\cal R}_\nu$ (which is modified by the $\sigma$
parameter through many factors, e.g. ${\cal R}_B$, $f_a$, $E_K$ and
$t_{dec}$, see eq.[\ref{Rnu}]).

\subsection{Sample lightcurves\label{sec:lc}}

We now calculate the typical early optical afterglow lightcurves for
various parameter regimes. Equation (\ref{Econservation}) states that
the initial afterglow energy, which is essentially the kinetic part of
the total energy, decreases with $\sigma$ given a constant total energy
$E=E_K+E_P$. At high-$\sigma$, not only the reverse shock flux level
drops, the forward shock flux level shortly after the shock crossing
also decreases steadily. At later times, the forward shock level would
increase due to magnetic energy injection. Since we are focusing on
the early afterglow emission, this effect will be neglected in the
following discussions. To explore the $\sigma$-effect, we fix the total
energy of the fireball so that $E_K$ decreases with increasing
$\sigma$. To simplify the calculations, we assume
${\cal R}_e \sim 1$ and $y \sim 1$.
The input parameters we adopt include $E_{52}=1$, $\gamma_0=150$, $n=1$,
$\epsilon_{e,f}=0.1$, $\epsilon_{B,f}=0.001$, $p_f=2.2$ and $z=1$
(with the standard cosmological parameters 
$\Omega_\Lambda \sim 0.7$, $\Omega_m \sim 0.3$ and $H_0 \sim 70~{\rm
km~s^{-1}~{Mpc}^{-1}}$). This gives the forward 
shock peak time and flux (Sari, Piran \& Narayan 1998; Kobayashi \&
Zhang 2003a) 
\ba
t_{p,f} & \sim & 1000 ~\mbox{s} \\
F_{p,f} & \sim & 1.7 (1+\sigma)^{-1}  ~\mbox{mJy} \nonumber \\
& & [m_R \sim 15.6
+2.5 \log (1+\sigma)]~.
\label{Fpf}
\ea
We also have $t_\gamma = [(3E/4\pi\gamma_0^2 n m_p
c^2)^{1/3}/2\gamma_0^2 c](1+z) \sim 60$ s. (For this calculation, we
have added in all the precise coefficients previously neglected.) We
then take two typical 
values of GRB durations. For the first case, we take $T=100$ s. When 
$\sigma$ is small the burst is in the thick shell regime (region
I). As $\sigma$ increases, the burst is in the non-spreading thin
shell regime (region II). For the second case, we take $T=20$ s. The
burst is always in the thin shell regime for any $\sigma$ value, but
transform from the spreading thin shell regime (region III) to the
non-spreading thin shell regime (region II) when $\sigma$ is large
enough. 

\begin{figure}
\epsscale{.80}
\plotone{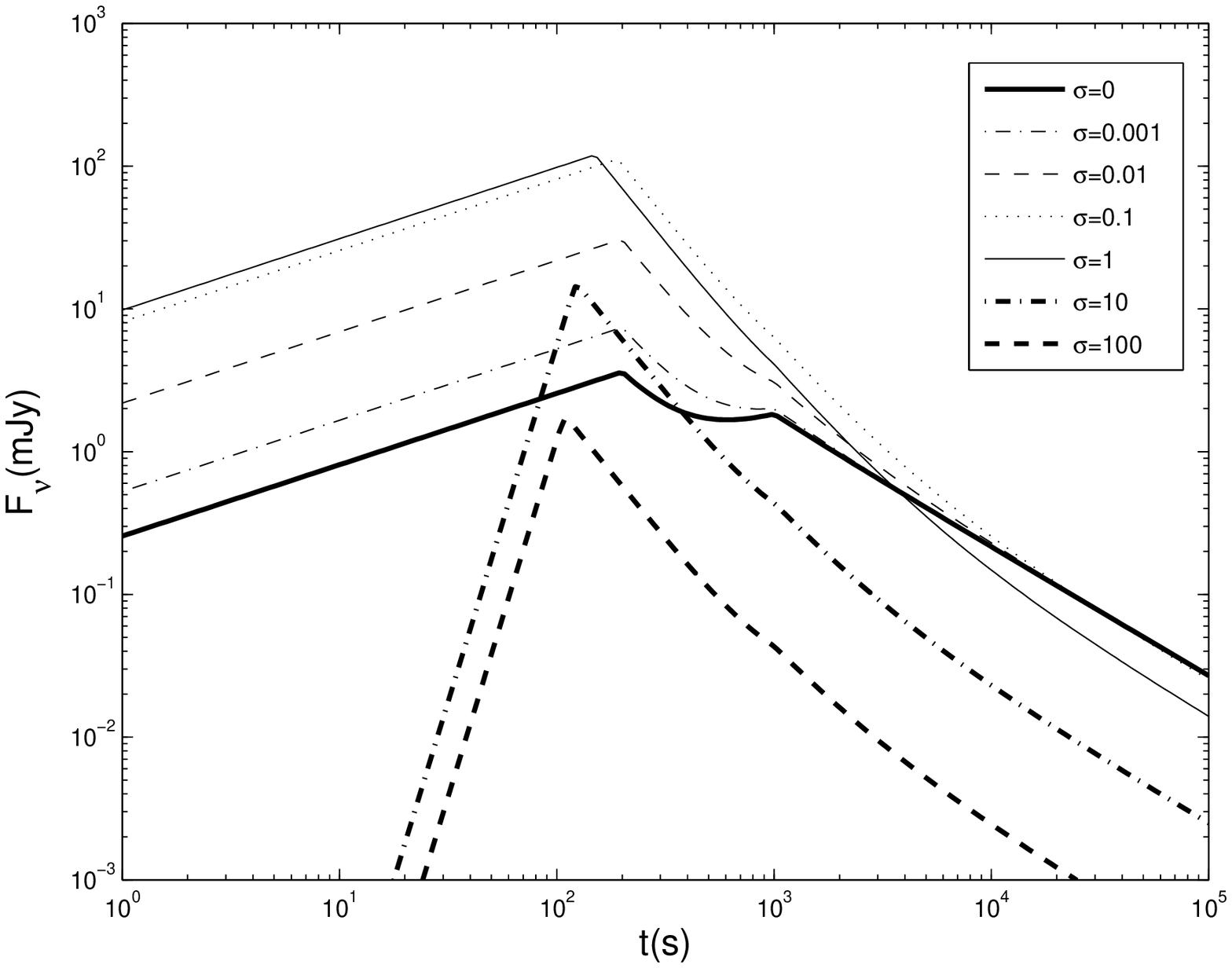}
\plotone{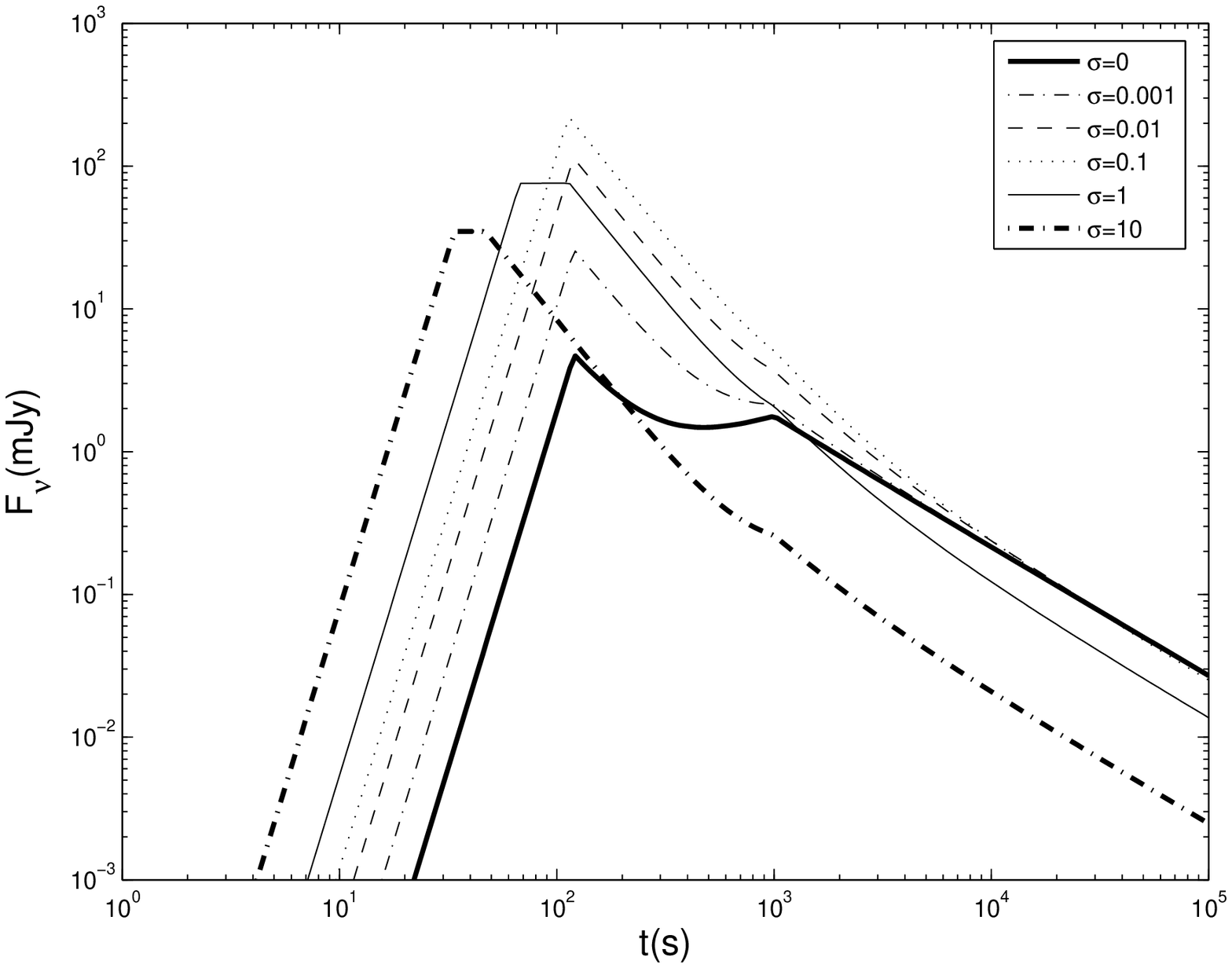}
\caption{
Sample early afterglow lightcurves for GRBs with an arbitrary
magnetization parameter $\sigma$. Following parameters are adopted.
$E_{52}=1$, $\gamma_0=150$, $n=1$, $\epsilon_{e,f}=0.1$,
$\epsilon_{B,f}=0.001$, $p_f=2.2$, and $z=1$. We also take ${\cal R}_e
\sim 1$ and $y \sim 1$. Both the forward shock and the
reverse shock emission components are calculated and they are
superposed to get the final lightcurve. For the forward shock emission
component we assumed that the time scale for the energy transfer from
the Poynting energy to the afterglow energy is long enough, so that
the forward shock level is defined by $E_K$ only, and its level
decreases with $\sigma$. This approximation is good shortly after the
reverse shock crossing. At later times, the real level could be
progressively higher than this level and the calculation should be
regarded as a lower limit. Lightcurves are calculated for
different $\sigma$ values. Thick solid: $\sigma=0$; thin dash-dotted:
$\sigma=0.001$; thin dashed: 
$\sigma=0.01$; thin dotted: $\sigma=0.1$; thin solid: $\sigma=1$;
thick dash-dotted: $\sigma=10$; thick dashed: $\sigma=100$.
$\gamma_{34}=1.5$ has been assumed for thin shell regimes. For
high-$\sigma$ cases, $\gamma_{34}$ is closer to unity, and the reverse
shock peak flux should be further suppressed.
(a) $T=100$ s case. According to eq.(\ref{RS-condition-2}), the
reverse shock exists when $\sigma < 200$. (b) $T=20$ s case. The
condition for the existence of the reverse shock is $\sigma < 20$.
\label{fig:lc}
}
\end{figure}

For each $T$ value, we calculate both the reverse shock and the
forward shock lightcurve for
several values of $\sigma$, i.e. $\sigma=0,0.001, 0.01, 0.1, 1, 10,
100$ (Fig.\ref{fig:lc}), as long as the condition for the existence of
the reverse shock (eq.[\ref{RS-condition-2}]) is satisfied.
The procedure of our calculation is the following. First, 
with $t_\gamma$, $T$ and the assumed $\sigma$, one can judge which
parameter region the burst is in. With this information one can then
calculate $t_{dec}=t_{p,r}$ and ${\cal R}_t$ for both the thick and
thin shell regimes, as well as $t_\times$
for the thin shell case. Next, we calculate ${\cal R}_B$ 
with the assumed $\sigma$ value (Fig.\ref{fig:epsB}). For the thick
shell case, we use the $\bar\epsilon_B$ value for $\gamma_{34} \sim
1000$ since $\bar\epsilon_B$ is insensitive to $\gamma_{34}$ when it
is large. For the thin shell case we use the $\bar\epsilon_B$ value
for $\gamma_{34} \sim 1.5$, exclusively\footnote{Notice that in
reality, when $\sigma$ is very large, $\gamma_{34}$ could be much
closer to unity. In such cases, the reverse shock peak flux should be
further suppressed.}. One can then solve ${\cal
R}_\nu$ (eq.[\ref{Rt2}]), and then use the value of ${\cal R}_\nu$
to calculate ${\cal R}_F$ (eq.[\ref{RF2}]), and hence
$F_{\nu,p,r}$. Since we know the 
temporal indices of the reverse shock lightcurve during the rising
($\sim 1/2$ for thick shell and $\sim 5$ for thin shell, Kobayashi
2000) and the decaying phase ($\sim -1.9$ for $p_r=2.2$), the reverse
shock lightcurve can be calculated once $t_{p,r}=t_{dec}$ and
$F_{\nu,p,r}$ are known. For the thin shell case, with the current
approximation, we roughly keep $F_\nu$ a constant between
$t_\times$ to $t_{dec}$, both of which are known. For the 
forward shock emission, the temporal index is $3(1-p_f)/4$ ($\sim 0.9$
for $p_f=2.2$) after the peak time, and is 1/2 before the peak time
(but after the deceleration time)\footnote{Before the
deceleration time, the forward shock lightcurve should have 
different temporal slopes. During the shock crossing, we have $\gamma_2 
\propto t^0$ for thin shells and $\gamma_2 \propto t^{-1/4}$ for thick 
shells. Using the standard synchrotron radiation analysis (e.g. Sari
et al. 1998), the forward shock emission temporal slope is 3 and 4/3
for the thin and thick shell cases, respectively. Between the shock
crossing time and the deceleration time in the high-$\sigma$ thin 
shell case, the temporal slope is flat.}. Given $F_{p,f}$ (which
is dependent on $\sigma$ (eq.[\ref{Fpf}]), the forward shock
lightcurve is also calculated.

Some sample R-band early afterglow lightcurves are presented in
Fig.\ref{fig:lc}, with the contributions from both the reverse and the
forward shocks superposed. For the forward shock emission
component we assume that the time scale for the energy transfer from
the Poynting energy to the afterglow energy is long enough, so that
the forward shock level is defined by $E_K$ only, and its level
decreases with $\sigma$. This approximation is good shortly after the
reverse shock crossing. At later times, the real level could be
progressively higher than this level and the calculation should be
regarded as a lower limit (see Zhang \& Kobayashi 2005 for more
explanations). In Fig.\ref{fig:lc}a, the cases for $T=100$
s are calculated. According to eq.(\ref{RS-condition-2}), a reverse
shock exists as long as $\sigma < 200$. We therefore calculate the
lightcurves up to $\sigma=100$. 
We can see that for $\sigma \leq 1$, the parameters
are in the thick shell (relativistic reverse shock) regime. When
$\sigma$ increases from below, the contrast between the reverse and
forward shock peak fluxes (i.e. ${\cal R}_F$) increases
steadily. Since the forward shock emission level does not change much
when $\sigma < 1$, the reverse shock peak flux increases steadily with
$\sigma$. At even higher $\sigma$ values, the reverse shock peak flux
drops steadily with 
$\sigma$. In the mean time, the burst enters the thin shell regime so
that the separation between $t_\times$ and
$t_{dec}$ becomes wider, and the reverse shock emission has a 
broader peak. In Fig.\ref{fig:lc}b, the cases
for $T=20$ s are calculated. According to eq.(\ref{RS-condition-2}), a
reverse shock exists as long as $\sigma < 20$, and we calculate the
lightcurves up to $\sigma=10$. The shell is in the thin shell regime
throughout the whole $\sigma$ range calculated. The transition from
spreading to non-spreading thin shell regime does not bring any
noticeable signature in the lightcurves. Again,  
the reverse shock peak flux increases with $\sigma$ initially (when
$\sigma \leq 0.1$), and starts to decrease when $\sigma \geq 0.1$. The
reverse shock peak is  
broad, but the separation gradually shrinks due to the decrease of the
$(Q^{-1/2}-Q^{-1})$ parameter (eq.[\ref{width}]). Throughout our
calculations, ${\cal R}_\nu$ remains larger 
than 25 (up to $\sim 1000$ for $\sigma=0$ in the $T=100$ s case), so
that our treatment by neglecting ${\cal R}_\nu < 1$ regime is
justified. 

We notice several interesting features from our results. First, the
reverse shock component is still noticeable even with $\sigma \simg
1$ (until reaching several tens or even hundreds when the condition
(\ref{RS-condition-2}) is no longer satisfied).
The absolute reverse shock peak flux increases with $\sigma$
initially, but drops steadily when $\sigma > 1$. 
Second, the forward shock emission level right after shock crossing
also drops with $\sigma$. This is because only the kinetic energy of
the baryonic component ($E_K$) defines the afterglow level after the
shock crossing time. The forward shock level will increase later due
to the transfer of the remaining magnetic energy into the
medium. One then expects an initially dim
early afterglow for a high-$\sigma$ flow, which would be brightened at
later times. If GRB prompt emission is due to magnetic dissipation
(e.g. Drenkhahn \& Spruit 2002), and if $\sigma$ is still
high in the afterglow phase (e.g. $\sim 10$), one may account for
the very large apparent GRB efficiencies inferred from some GRBs
(e.g. Lloyd-Ronning \& Zhang 2004). Such a picture may be also
relevant to the recent December 27 giant flare
afterglow from the soft gamma-ray repeater 1806-20, for which a very
high gamma-ray efficiency is inferred (Wang et al. 2005
and references therein). Third, the broad reverse shock 
peak is a novel feature identified in the high-$\sigma$ model, it can
be used to diagnose the existence of a Poynting-flux-dominated
flow. The physical origin of the broad peak is that a high-$\sigma$
value leads the decoupling of the shock crossing radius $R_\times$ and
the deceleration radius $R_\gamma$, so that multi-crossing of a series
of successive shocks leads to continuous heating of 
the ejecta shell before cooling starts.

In the above calculations, $y=1$ has been adopted (i.e. we assume that 
the pair fraction is negligible in the ejecta). In some cases,
especially in the high-$\sigma$ regime, $y$ could be much larger than
unity. It would be essential to investigate the $y$-dependence of the
current analysis. Solve ${\cal R}_\nu$ from eq.(\ref{Rt2}) and submit
it to ${\cal R}_F$, we find ${\cal R}_F \propto
y^{(7-4\alpha)/3}$, which is $\propto y^{-1/3}$ for
$\alpha=2$. We can see that a larger $y$ will lower the
reverse-to-forward shock peak flux contrast, although the dependence
is mild. The ${\cal R}_F$ factor is more sensitive to ${\cal R}_e$ (i.e.
$\propto {\cal R}_e^{4(\alpha-1)/3}$), but assuming a similar shock
acceleration mechanism, ${\cal R}_e$ may not deviate too much from
unity. 

Our results can be directly compared with the early afterglow data
of the four bursts whose such information is available so far. The
case of GRB 990123 (Akerlof et al. 1999) is consistent with a flow
with $0.1<\sigma<1$ in which regime ${\cal R}_F$ is large and the
reverse shock peak is not broadened. The observed bright afterglow
also argues against a higher-$\sigma$ flow. The case of GRB 021211
(Fox et al. 2002b; Li et al. 2002a) also shows a large ${\cal R}_F$,
which also suggests that $\sigma > 0.1$. For GRB 021004,
Kobayashi \& Zhang (2003a) attempted to fit the data with the
$\sigma=0$, ${\cal R}_B=1$ model. Another data point at an earlier
epoch after the burst trigger reported by Fox et al. (2003a) makes
that model more difficult to fit, and it has been attributed to a
continuously energy injection (Fox et al. 2003a) or to the emission
from a wind-type medium (Li \& Chevalier 2003). However, using the
theory developed in this paper, the data may be consistent with a
high-$\sigma$ flow (e.g. $\sigma \simg 10$), so that the extended
early afterglow emission could be interpreted as the combination of
the broad reverse shock peak and the gradual transfer of the Poynting
energy into the afterglow energy. In the high-$\sigma$ regime, the
$(\gamma_{34}-1)$ is quite small during the shock crossing, which will
lower the reverse shock peak flux and ${\cal R}_F$.
Similarly, a broad early afterglow bump was identified in GRB
030418 (Rykoff et al. 2004) which challenges the conventional reverse
shock model but may be consistent with a high-$\sigma$ flow (weak or
no reverse shock component).
Although detailed modeling is needed (we plan to do it in a future
work), we tentatively conclude that all the current early optical
afterglow data may be understood within the theoretical framework
developed in this paper, if $\sigma$ is allowed to vary for different
GRB fireballs.

\section{Conclusions and discussion}

We have derived a rigorous analytical solution for the relativistic
90$^{\rm o}$ shocks under the ideal MHD condition (eq.[\ref{eq}]).
Generally, the solution depends both on the magnetization $\sigma$
parameter and the Lorentz factor of the shock, $\gamma_{12}$. The
solution can be reduced to the Blandford-McKee hydrodynamical solution
when $\sigma=0$, and to the Kennel-Coroniti solution (which depends on
$\sigma$ only) when the $\gamma_{21} \gg 1$. Our generalized solution
can be used to treat the more general cases, e.g. when the reverse
shock upstream and downstream are mildly relativistic with each
other. Since GRBs invoke a shell with finite 
width, this latter possibility is common (e.g. the parameter space for
thin shell greatly increases in the high-$\sigma$ regime), so that our
generalized solution is essential to deal with the GRB reverse shock
problem.

Several interesting conclusions emerge from our analysis. (1) Strong
shocks still exist in the high-$\sigma$ regime, as long as the shock
is relativistic. Figs.\ref{fig:sig}c,d indicate that as $\sigma$
increases, both the downstream ``temperature'' $e_2/n_2$ and the
``shock compression factor'' $n_2/n_1$ decreases with respect to the
$\sigma=0$ values. However, the suppression factors in both cases are
only mild (a factor of $\sim$ 0.5), and they saturate when $\sigma \gg
1$. In the relativistic shock 
regime, the results are actually consistent with Kennel \& Coroniti
(1984). However, these authors did not calculate the suppression factor
with respect to the $\sigma=0$ case, and did not explore further into the
high-$\sigma$ regime, so that their results leave the impression to
the readers that the shock is completely suppressed when $\sigma$
reaches higher values. For typical GRB parameters, we found that a
reverse shock still exists when $\sigma$ is as high as several tens or
even hundreds. When the reverse shock exists, its emission level
decreases when $\sigma$ gets higher. This is not only because
the reverse shock becomes weaker since $\gamma_{34}$ gets close to
unity in the high-$\sigma$ regime, but also because
the total kinetic energy in the flow (which is the energy reservoir
for shock dissipation) gets smaller given a same total energy.
(2) When discussing ejecta-medium interaction,
somewhat surprisingly, some important parameters, such as $F$ and
$C_\Delta$, are very insensitive to the reverse shock Lorentz factor,
$\gamma_{34}$, and can be regarded as the function of $\sigma$ only
(\S\ref{sec:general}, \S\ref{sec:regimes}). This 
greatly simplify the problem, and is essential to characterize the
parameter regimes. (3) The triple coincidence of the first three
critical lines in eq.(\ref{lines}) is very crucial for a
self-consistent description of the problem (\S\ref{sec:regimes}).

Comparing with the conventional hydrodynamical treatment, we reveal
several novel features for the early lightcurves. First, as $\sigma$
increases, the reverse shock peak flux increases
rapidly initially, reaching a peak around $\sigma \sim (0.1-1)$, and
starts to decrease when $\sigma \geq 1$. 
Second, due to the inability of tapping the Poynting flux energy
during the shock crossing process, the fireball deceleration radius
for the thin shell case decreases as $\sigma$ increases $[\propto
(1+\sigma)^{-1/3}]$. 
The forward shock emission level is also lower right after shock
crossing. Third, in the high-$\sigma$ thin shell 
regime, the reverse shock peak is broadened due to the separation of
the shock crossing radius and the deceleration radius. This is a
signature for a high-$\sigma$ flow, which can be used to diagnose the
magnetic content of the fireball. Fourth, as $\sigma$ becomes large
enough (larger than several tens or several hundreds), the condition
for forming a reverse shock is no longer satisfied, and there should
be no reverse shock component in the early afterglow lightcurves. This
could be consistent with very early tight optical upper limits
for some GRBs, such as the recent Swift dark burst GRB 050319a (Roming
et al. 2005). In summary, the above new features allows the
current theory to potentially interpret known GRB early
afterglow cases collected so far as well as the case of the dark
bursts, if one allows $\sigma$ to vary in a wide enough range (say,
from $0.01$ to $100$). The Swift GRB mission, launched on November 20,
2004, is expected to 
detect many early optical afterglow lightcurves with the UV-optical
telescope on board. We expect to further test our theoretical
predictions against the abundant Swift data, and to systematically
diagnose the magnetic content of GRB fireballs. 

If the GRB ejecta is indeed magnetized, as inferred from the early
afterglow data, the internal shocks should also be corrected by the
magnetic suppression factor. This aspect has been investigated by Fan,
Wei \& Zhang (2004b) recently.

Throughout the paper, we have treated the problem under the ideal MHD
limit. In the high-$\sigma$ case, strong magnetic dissipation may
occur. The magnetic dissipation effect has been included in the
internal shock study of Fan et al. (2004b). Our treatment in this
paper presents a first order picture to the early afterglow problem
(see also Fan et al. 2004a, whose treatment in the mildly magnetized
regime is consistent with ours), and further considerations are needed
to fully delineate the physics involved. 
Also our whole discussion is relevant when a reverse shock is
present. It does not apply to the regime for an even higher $\sigma$
value (e.g. Lyutikov \& Blandford 2003).
Finally, as discussed in \S4.1, further studies on particle
acceleration in MHD shocks are essential to give a more accurate
calculation on the reverse shock emission in the high-$\sigma$
regime. 

We only discussed one type of the medium, i.e., assuming a constant
medium density, typically for the interstellar medium. In principle,
the medium density can vary with distance from the central engine. In
particular, a wind-type medium, characterized by the $n \propto
R^{-2}$ profile, has been widely discussed. Our MHD shock theory could
be straightforwardly used to the wind case to study the reverse shock
emission in combination with the previous pure hydrodynamical
treatments (Chevalier \& Li 2000; Wu et al. 2003; Kobayashi \& Zhang
2003b; Kobayashi, \Mesz~\& Zhang 2004, see Fan et al. 2004a for a
preliminary treatment).

\acknowledgments

We thank the anonymous referee for a thorough check of the scientific
content of the paper, Y. Z. Fan and M. Lyutikov for important remarks
and in-depth discussions, P. \Mesz, L. J. Gou, and D. M. Wei for
helpful comments, and C. Akerloff, M. G. Baring, R. D. Blandford,
D. Fox, C. Kouveliotou, P. Kumar, J. Granot, E. Nakar, T. Piran,
R. Sari, A. Spitkovsky, M. J. Rees, E. Rykoff, and  W. Zhang for
discussions. This work is supported by NASA grants NNG04GD51G and
NAG5-13286 (for B.Z.), by Penn State Center for Gravitational Wave
Physics through NSF PHY 01-14375 and by Eberly Research Fund of Penn
State University (for S.K.) and by a Swift GI (Cycle 1) theory program
(for both authors).

\appendix

\section{Derivation of the solution of the relativistic 90$^{\rm o}$ shock
jump conditions}

\subsection{Lorentz transformations}

Given the definitions of $\gamma_{ij}$, $\beta_{ij}$ and $u_{ij}$,
the following Lorentz transformations are frequently used in the
derivations.
\begin{eqnarray}
\beta_{2s} & = & \frac{\beta_{1s}-\beta_{21}}{1-\beta_{1s}
\beta_{21}} \\
\beta_{1s} & = & \frac{\beta_{2s}+\beta_{21}}{1+\beta_{2s}
\beta_{21}} \\
\beta_{21} & = & \frac{\beta_{1s}-\beta_{2s}}{1-\beta_{1s}
\beta_{2s}} \\
\gamma_{2s} & = & \gamma_{1s} \gamma_{21} (1-\beta_{1s}\beta_{21}) \\
\gamma_{1s} & = & \gamma_{2s} \gamma_{21} (1+\beta_{2s}\beta_{21}) \\ 
\gamma_{21} & = & \gamma_{1s} \gamma_{2s} (1-\beta_{1s}\beta_{2s}) \\
u_{2s} & = & \gamma_{1s}\gamma_{21} (\beta_{1s}-\beta_{21}) \\
u_{1s} & = & \gamma_{2s}\gamma_{21} (\beta_{2s}+\beta_{21}) 
\label{LT} \\ 
u_{21} & = & \gamma_{1s}\gamma_{2s} (\beta_{1s}-\beta_{2s}) \\
\beta_{1s}-\beta_{2s} & = & \frac{u_{21}}{\gamma_{1s}\gamma_{2s}}
\end{eqnarray}

\subsection{Derivation of equation (\ref{e/n})}

Let us define (Kennel \& Coroniti 1984)
\begin{equation}
Y \equiv \frac{B_{2s}}{B_{1s}}=\frac{\gamma_{2s}u_{1s}}
{\gamma_{1s}u_{2s}} = \frac{\beta_{1s}}{\beta_{2s}},
\end{equation}
the equations (\ref{jump2}) and (\ref{jump3}) can be re-written as
\begin{eqnarray}
\gamma_{1s}\mu_1[1+(1-Y)\sigma] & = & \gamma_{2s} \mu_2
\label{jump2b}\\
u_{1s}\mu_1\left[ 1+\frac{\sigma}{2\beta_{1s}^2} (1-Y^2)\right]
 & = & u_{2s}\mu_2 + \frac{p_2}{n_2 u_{2s}}. \label{jump3b}
\end{eqnarray}
Multiply eq.(\ref{jump2b}) by $\beta_{1s}$ and substitute the
resultant formula into eq.(\ref{jump3b}), one can derive
\begin{eqnarray}
\frac{n_2 m_p c^2+e_2}{n_2} & = & \left(\gamma_{1s}\gamma_{2s}
[1+(1-Y)\sigma] - u_{1s}u_{2s}[1+\frac{\sigma}{2} (\beta_{1s}^{-2}
- \beta_{2s}^{-2})]\right) \mu_1 \nonumber \\
& = & \left(\gamma_{21} - \frac{u_{21}^2}{2
u_{1s}u_{2s}}\sigma\right) \mu_1.
\end{eqnarray}
The equation (\ref{e/n}) in the text can be then derived
straightforwardly. 

\subsection{Solving $u_{2s}^2$}
Combining eq.(\ref{jump2b}) and the definition of $\mu_2$
(eq.[\ref{mu}]), one can derive
\begin{equation}
\frac{\gamma_{1s}}{\gamma_{2s}}[1+(1-Y)\sigma] = 1+\hat\Gamma
(\gamma_{21}-1)-\hat\Gamma \frac{u_{21}^2}{2u_{1s}u_{2s}}\sigma.
\end{equation}
This turns out to be a three-order equation
of $x\equiv u_{2s}^2$, i.e. 
\begin{equation}
Ax^3+Bx^2+Cx+D=0,
\label{eq}
\end{equation}
where
\begin{eqnarray}
A & = & \hat\Gamma(2-\hat\Gamma)(\gamma_{21}-1)+2 \\
B & = & -(\gamma_{21}+1) \left[(2-\hat\Gamma)(\hat\Gamma
\gamma_{21}^2+1)+\hat\Gamma(\hat\Gamma-1)\gamma_{21}\right]\sigma
\nonumber \\
& & -(\gamma_{21}-1)\left[\hat\Gamma(2-\hat\Gamma)
(\gamma_{21}^2-2)+(2\gamma_{21}+3)\right] \\
C & = & (\gamma_{21}+1) \left[\hat\Gamma(1-\frac{\hat\Gamma}{4})
(\gamma_{21}^2-1)+1 \right] \sigma^2 \nonumber \\
& & + (\gamma_{21}^2-1) \left[2\gamma_{21}-(2-\hat\Gamma)
(\hat\Gamma \gamma_{21}-1)\right] \sigma \nonumber \\
& & + (\gamma_{21}+1)(\gamma_{21}-1)^2(\hat\Gamma-1)^2 \\
D & = & -(\gamma_{21}-1)(\gamma_{21}+1)^2 (2-\hat\Gamma)^2
\frac{\sigma^2}{4}~.
\end{eqnarray}
For $\hat\Gamma=4/3$, the four coefficients could be written
equivalently as
\begin{eqnarray}
A & = & 8\gamma_{21}+10 \\
B & = & -(\gamma_{21}+1)(8\gamma_{21}^2+4\gamma_{21}+6)\sigma
-(\gamma_{21}-1)(8\gamma_{21}^2+18\gamma_{21}+11) \\
C & = & (\gamma_{21}+1)(8\gamma_{21}^2+1)\sigma^2
+(\gamma_{21}^2-1)(10\gamma_{21}+6)\sigma
+(\gamma_{21}+1)(\gamma_{21}-1)^2 \\
D & = & -(\gamma_{21}-1)(\gamma_{21}+1)^2 \sigma^2~.
\end{eqnarray}

\subsubsection{$\sigma=0$ limit\label{sec:append1}}

When $\sigma=0$, the equation (\ref{eq}) is reduced to
\be
x\left[x-(\gamma_{21}^2-1)\right]
\left\{[\hat\Gamma(2-\hat\Gamma)(\gamma_{21}-1)+2]x
-(\gamma_{21}-1)(\hat\Gamma-1)^2 \right\}=0~,
\label{eq1}
\ee
which gives equation (\ref{bm0}) besides the other two non-physical
solutions $u_{2s}=0$ and $u_{2s}=u_{21}$.

\subsubsection{$\gamma_{21} \gg 1$ limit\label{sec:append2}}

When $\gamma_{21} \gg 1$, the $x^3$ term is a small quantity and
is negligible. The equation (\ref{eq}) is reduced to
\be
\hat\Gamma(2-\hat\Gamma)(\sigma+1)x^2
-\left[\hat\Gamma(1-\frac{\hat\Gamma}{4})\sigma^2
+(\hat\Gamma^2 - 2\hat\Gamma+2)\sigma+(\hat\Gamma-1)^2\right] x
+(2-\hat\Gamma)^2 \frac{\sigma^2}{4} =0~.
\label{eq2}
\ee
This gives the solution (\ref{u2skc}) in the text (when the
non-physical solution is neglected).


\section{Notation list}

\begin{tabular}{ll}
subscript 1 & upstream (\S2 and Appendix A), or unshocked medium (\S3
and \S4) \\
subscript 2 & downstream (\S2 and Appendix A), or shocked medium (\S3
and \S4) \\
subscript 3 & shocked ejecta \\
subscript 4 & unshocked ejecta \\
subscript $s$ & shock \\
$c$ & speed of light \\
$e_i$ & internal energy density in region $i(=1,2,3,4)$ \\
$f_a$ & correction factor of $(e_2/n_2m_pc^2)$ normalized to the
$\sigma=0$ value \\
$f_b$ & correction factor of $(n_2/n_1)$ normalized to the
$\sigma=0$ value \\
$f_c$ & magnetic-to-thermal pressure ratio plus 1 \\
$g_f$ & $(p_f-2)/(p_f-1)$ \\
$g_r$ & $(p_r-2)/(p_r-1)$ \\
$l$ & Sedov length \\
$m_e$ & electron rest mass \\
$m_p$ & proton rest mass \\
$n_i$ & baryon number density in region $i(=1,2,3,4)$ \\
$p_i$ & thermal pressure in region $i(=1,2,3,4)$ \\
$p_{b,i}$ & magnetic pressure in region $i(=1,2,3,4)$ \\
$p_f$ & electron power law index in the forward shock \\
$p_r$ & electron power law index in the reverse shock \\
$t_{dec}$ & deceleration time measured by the observer \\
$t_\gamma$ & $R_\gamma/C_\gamma \gamma_4^2 c$ (eq.[\ref{tgam}]) \\
$t_{p,f}$ & emission peak time of the forward shock component \\
$t_{p,r}$ & emission peak time of the reverse shock component \\
$t_\times$ & shock crossing time measured by the observer \\
$u_{ij}$ & four speed in the region $i(=1,2,3,4)$ in the rest frame of 
$j(=1,2,3,4,s)$ \\
$x$ & $u_{2s}^2$ \\
$y$ & pair multiplicity parameter \\
$z$ & redshift 
\end{tabular}

\begin{tabular}{ll}
$A$, $B$, $C$, $D$ & coefficients to solve the equation for
$u_{2s}^2$ \\
$B_i$ & comoving magnetic field in the region $i(=1,2,3,4)$ \\
$B_{is}$ & magnetic field in the region $i(=1,2,3,4)$ in the rest
frame of the shock \\
$B_f$ & comoving magnetic field in the forward shocked region \\
$B_r$ & comoving magnetic field in the reverse shocked region \\
$C_N$ & correction factor to $R_N$ with respect to the $\sigma=0$ case 
\\
$C_\Delta$ & correction factor to $R_\Delta$ with respect to the
$\sigma=0$ case \\
$C_\gamma$ & correction factor to $R_\gamma$ with respect to the
$\sigma=0$ case \\
${\cal E}$ & shock frame electric field \\
$E$ & isotropic total energy of the fireball \\
$E_K$ & isotropic kinetic energy of the fireball \\
$E_P$ & isotropic Poynting-flux energy of the fireball \\
$F$ & the product of $f_a$, $f_b$ and $f_c$ \\
$F_\nu$ & specific flux at the frequency $\nu$ \\
$F_{\nu,m,f}$ & maximum synchrotron emission specific flux in the
forward shock \\
$F_{\nu,m,r}$ & maximum synchrotron emission specific flux in the
reverse shock \\
$F_{\nu,p,f}$ & peak flux for the forward shock emission
component in certain (e.g. R) band \\
$F_{\nu,p,r}$ & peak flux for the reverse shock emission
component in certain (e.g. R) band \\
$H_0$ & Hubble constant \\
$M_0$ & mass in the ejecta \\
$M_{\rm ISM}$ & mass of the interstellar medium collected by the shock \\
$N_b$ & total baryon number in the shell \\
$N_\pm$ & total electron-positron pair number in the shell \\
$N_{e,f}$ & lepton (electron) number in the forward shock \\
$N_{e,r}$ & lepton (electron and pairs) number in the reverse shock \\
$Q$ & a parameter introduced to categorize the parameter regimes
(defined by eq.[\ref{Q}])\\
$R$ & radius from the central engine \\
${\cal R}_B$ & reverse-to-forward comoving magnetic field ratio \\
${\cal R}_e$ & reverse-to-forward ratio of the $\epsilon_e g$
parameter \\
${\cal R}_F$ & reverse-to-forward peak flux ratio \\
${\cal R}_t$ & forward-to-reverse peak time ratio \\
${\cal R}_\nu$ & the ratio between $\nu_R$ and $\nu_{m,r}(t_{dec})$ \\
$R_N$ & radius where the reverse shock becomes relativistic \\
$R_s$ & radius where the ejecta shell starts to spread \\
$R_\Delta$ & radius where the reverse shock crosses the ejecta shell
\\
$R_\gamma$ & radius where the fireball collects $1/\gamma_0$
rest mass of the fireball \\
$T$ & central engine activity time scale \\
$U_{\rm B,0}$ & initial comoving magnetic energy \\
$U_{\rm B}$ & comoving magnetic energy after shock crossing \\
$X$ & an intermediate parameter introduced in eq.(\ref{u2skc}) \\
$Y$ & ratio between $B_{2s}$ and $B_{1s}$ 
\end{tabular}

\begin{tabular}{ll}
$\hat\Gamma$ & adiabatic index \\
$\Delta$ & shell width in the lab frame \\
$\Delta_0$ & initial shell width in the lab frame \\
$\Omega_m$ & cosmology mass density parameter \\
$\Omega_\Lambda$ & cosmology $\Lambda$ density parameter \\
$\alpha$ & temporal decay index of the reverse shock emission
component after peak time \\
$\beta_{ij}$ & dimensionless velocity of region $i(=1,2,3,4)$ in the
rest frame of $j(=1,2,3,4,s)$ \\
$\hat\gamma$ & an equivalent Lorentz factor defined in
eq.(\ref{hatgam}) \\
$\gamma_{ij}$ & Lorentz factor of region $i(=1,2,3,4)$ in the
rest frame of $j(=1,2,3,4,s)$ \\
$\gamma_i$ & Lorentz factor of region $i(=2,3,4)$ in the rest frame of 
the circumburst medium \\
$\gamma_0$ & initial Lorentz factor of the fireball, $\gamma_0 \equiv
\gamma_4$ \\
$\gamma_c$ & critical initial Lorentz factor that separates thick
vs. thin shell regimes (eq.[\ref{gammac}]) \\
$\gamma_\times$ & fireball Lorentz factor at the shock crossing time \\
$\gamma_{e,m,f}$ & electron minimum Lorentz factor in the forward
shock \\
$\gamma_{e,m,r}$ & electron minimum Lorentz factor in the reverse
shock \\
$\epsilon_{e,f}$ & electron energy equipartition parameter in the
forward shock \\
$\epsilon_{e,r}$ & electron energy equipartition parameter in the
reverse shock \\
$\epsilon_{B,f}$ & magnetic energy equipartition parameter in the
forward shock \\
$\bar\epsilon_{B,r}$ & equivalent magnetic energy equipartition
parameter in the reverse shock \\
$\mu_i$ & specific enthalpy in region $i(=1,2,3,4)$ \\
$\nu_{c,f}$ & forward shock synchrotron cooling frequency \\
$\nu_{c,r}$ & reverse shock synchrotron cooling frequency \\
$\nu_{m,f}$ & forward shock synchrotron typical frequency \\
$\nu_{m,r}$ & reverse shock synchrotron typical frequency \\
$\nu_R$ & R-band frequency \\
$\xi$ & a parameter defined in eq.(\ref{xi}) \\
$\xi_0$ & the $\xi$ value when $\Delta=\Delta_0$ \\
$\sigma$ & magnetization parameter as defined in eq.(\ref{sigma}) \\
\end{tabular}

\end{document}